\documentclass[aps,prd,twocolumn,nofootinbib]{revtex4-2}
\usepackage{amssymb,amsmath}
\usepackage{epsfig,epsf}
\usepackage{bm} 
\usepackage{color} 
\usepackage{slashed}
\usepackage{relsize}	
\usepackage{soul} 
\usepackage{hyperref}
\DeclareMathOperator{\tr}{\mathrm{tr}}
\newcommand{\mean}[1]{\langle #1 \rangle}
\newcommand{\D}{{\rm d}}
\newcommand{\I}{{\rm i}}
\newcommand{\h}[1]{\widehat{#1}}
\newcommand{\E}{{\rm e}}
\newcommand{\de}{\partial}
\renewcommand{\vec}[1]{\ensuremath{\mathchoice				
		{\mbox{\boldmath$\displaystyle\mathbf{#1}$}}
		{\mbox{\boldmath$\textstyle\mathbf{#1}$}}
		{\mbox{\boldmath$\scriptstyle\mathbf{#1}$}}
		{\mbox{\boldmath$\scriptscriptstyle\mathbf{#1}$}}}}
\DeclareMathOperator*{\SumInt}{%
\mathchoice%
  {\ooalign{$\displaystyle\sum$\cr\hidewidth$\displaystyle\int$\hidewidth\cr}}
  {\ooalign{\raisebox{.14\height}{\scalebox{.7}{$\textstyle\sum$}}\cr\hidewidth$\textstyle\int$\hidewidth\cr}}
  {\ooalign{\raisebox{.2\height}{\scalebox{.6}{$\scriptstyle\sum$}}\cr$\scriptstyle\int$\cr}}
  {\ooalign{\raisebox{.2\height}{\scalebox{.6}{$\scriptstyle\sum$}}\cr$\scriptstyle\int$\cr}}
}
\begin{document}

\title{Chiral magnetic effect in a cylindrical domain}

\author{Matteo Buzzegoli}

\author{Kirill Tuchin}

\affiliation{
Department of Physics and Astronomy, Iowa State University, Ames, Iowa, 50011, USA}

\begin{abstract}
 We compute the chiral magnetic effect (CME) in a cylindrical region coaxial with the external magnetic field. As the boundary condition we require vanishing of the radial component of the electric current on the cylinder side wall. We find that when the magnetic length is comparable to or larger than the cylinder radius, the CME is suppressed compared to the corresponding result in an infinite medium. As a result, for a given cylinder radius, the suppression is stronger in weak fields. We  argue that the electric current generated by the CME vanishes at the cylinder wall and monotonically increases toward the symmetry axis. 
 
\end{abstract}

\maketitle

\section{Introduction}\label{sec:Intro}

Systems possessing chiral fermions can be found at vastly different energy and distance scales. For example, they include  the quark-gluon plasma, Dirac and Weyl semimetals and certain neutron stars. Owing to the chiral anomaly, these chiral systems exhibit remarkable new properties such as nondissipative transport
phenomena \cite{Kharzeev:2013ffa,Kharzeev:2012ph,Kharzeev:2015znc,Miransky:2015ava,Landsteiner:2016led,Kharzeev:2020jxw,Kharzeev:2007tn}. Since the chiral anomaly is an inherently quantum effect, the novel chiral properties associated with them are entirely absent in classical theories. Perhaps the most celebrated of them is the chiral magnetic effect (CME) \cite{Kharzeev:2004ey,Kharzeev:2007jp,Fukushima:2008xe} which refers to the induction of electric current along the direction of the external magnetic field:
\begin{equation}\label{cme}
\h{\vec{j}}= \frac{q^2}{2\pi^2} \mu_A \h{\vec{B}},
\end{equation}
where $\h{\vec{j}}$ is the electric current density, $\h{\vec{B}}$ is the magnetic field, $q$ is the electric charge, and $\mu_A$ is the axial chemical potential; the hat sign denotes operators. The coefficient in front of $\h{\vec{B}}$ is referred to as the CME conductivity.  The theoretical description of this effect was developed in many papers over the past decade. With a few notable exceptions, mentioned below, all studies ignore the boundary effects even though all realistic systems are confined in a finite region of space. A simple dimensional analysis can convince one of the importance of the finite-volume effects. The goal of this paper is to initiate study of the boundary conditions on the CME. To this end we examine the role of the cylindrical boundary whose symmetry axis coincides with the external magnetic field direction. The axial symmetry makes this model analytically tractable and allows us to develop keen insights into the CME dynamics in finite systems.

There are three reasons that necessitate imposing the boundary conditions. First, there are physical boundaries: nuclei, stars and material specimens all have finite dimensions. Second, many systems consist of domains of finite $\mu_A$ each generating different chiral magnetic currents. For instance, it was argued in \cite{Zhitnitsky:2012ej,Zhitnitsky:2013hs} that the quark-gluon plasma may contain one large chiral domain. The vanishing of $\mu_A$ outside this domain has a dramatic impact on CME \cite{Tuchin:2018rrw,Tuchin:2020nai}. Third, the spacetime geometry can confine a system to a finite volume as happens, for example, in a rapidly rotating chiral fluid. Such a system is realized in off-center relativistic heavy-ion collisions that produce fluid with extremely high vorticity \cite{STAR:2017ckg}, and the causality demands that
a particle cannot rotate with angular velocity $\Omega$ further than a distance of $R=c/\Omega$.
A number of studies have begun to examine the importance of the inclusion of these
casual boundaries \cite{Duffy:2002ss,Ambrus:2015lfr,Ebihara:2016fwa,Chernodub:2017ref,Chernodub:2017mvp,Chen:2017xrj,Sadooghi:2021upd,Ambrus:2023bid,Yang:2021hor}. Admittedly, a system dynamics may, and usually is, richer than can be encapsulated by a boundary condition. Nevertheless, modeling finite size with boundary conditions is a good starting point.  

The problem of consistent boundary conditions for a system of relativistic fermions has been addressed already at the dawn of QCD. The color confinement requires quarks to stay inside hadrons. In the MIT bag model this is accomplished by imposing boundary conditions at a spacelike surface \cite{Chodos:1974je,Thomas:1982kv,Shanker:1983yd}. Specifically, the normal component of the current is required to vanish. Many studies of the MIT bag model considered a spherical symmetry with a sphere as the boundary surface. We instead adjust the model to our problem by imposing the boundary condition on the side surface of a cylinder coaxial with the magnetic field. The height of the cylinder is immaterial so we simply assume that it extends to infinity. Although this model is axially symmetric, it breaks the chiral symmetry \cite{Chodos:1974je,Thomas:1982kv,Shanker:1983yd}. Indeed, 
the bouncing of a massless fermion off the boundary, flips its momentum  and helicity/chirality.

The dynamics in the transverse direction depends on two radial scales. One is the radius of the cylinder $R$, and the other one is the magnetic length $l_B=\hbar c /\sqrt{|qB|}$. The relative importance of the boundary condition on the properties of the system is indicated by the dimensionless parameter 
\begin{equation}\label{rhoR}
\rho_R=\frac{R^2}{2 l_B^2}.
\end{equation}
When this ratio is large $\rho_R\gg 1$, the boundary barely modifies the Landau states. However, when $\rho_R\sim 1$, $l_B$ is comparable to the radius of the cylinder and the finite size effects strongly deform the Landau levels and the properties of the system. 

The role of boundaries on the CME was discussed before in  \cite{Gorbar:2015wya,Valgushev:2015pjn,Sitenko:2016iqb} in the context of the two-dimensional slab geometry. When the magnetic field is directed along the normal to the flat slab boundary, analytical calculation yielded vanishing CME current \cite{Gorbar:2015wya,Sitenko:2016iqb}.
However, when the magnetic field is parallel to the boundaries of the slab, numerical lattice calculations showed that while the total current vanishes, it is locally finite near the slab boundary \cite{Valgushev:2015pjn}.
In our model, the axial chemical potential is introduced as a dynamical, albeit adiabatic, quantity and we preserve the axial symmetry of the magnetic field. Thus the generation of a CME current is not forbidden by the boundary conditions, which we confirm by explicit analytical calculation.

This paper is organized as follows. In Sec.~\ref{sec:Review} we review the derivation of the CME in thermal equilibrium at finite axial chemical potential using statistical mechanics. In Sec.~\ref{sec:DiracEq} we obtain the solutions of the
Dirac equation of a massless particle in a constant magnetic field using the MIT
boundary conditions in a finite cylinder. We show in Sec.~\ref{sec:Thermo} that
these solutions are compatible with a thermal equilibrium description.
Finally, in Sec.~\ref{sec:CME} we derive the CME in a finite cylinder. We will refer to it as the ``bound CME''in contrast to the ``unbound CME'' of the infinite system, given by (\ref{cme}). To compute the electric current along the magnetic field direction we use Vilenkin's method \cite{Vilenkin:1980fu} that consists in considering a system of relativistic fermions in thermal equilibrium at a finite axial chemical potential $\mu_A$, which describes the chiral imbalance.\footnote{ In these settings, the derivation must be carried out without including the effect of the electric field generated by the CME current;
otherwise, the resulting current would simply be vanishing~\cite{Vilenkin:1980ft}.} Although the CME is an out-of equilibrium phenomenon \cite{Yamamoto:2015fxa}, the use of thermal equilibrium is justified at timescales shorter than the relaxation time of the CME current. Our main results are Eq.~(\ref{eq:CMEAnalytic}) and Figs.~\ref{fig:NormCME_fixed_RhoR}--\ref{fig:CriticalB_fixed_R}. Our summary and outlook are presented in Sec.~\ref{sec:Conclusions}.
We use natural units where $\hbar=c=k_B=1$.

\section{The CME from statistical mechanics}
\label{sec:Review}
The CME can be computed using the methods of statistical mechanics assuming thermal equilibrium
at finite axial imbalance~\cite{Vilenkin:1980zv,Vilenkin:1980fu,Buzzegoli:2020ycf}.
Before addressing the effect of a finite boundary, in this section we review the
derivation of the CME for a system of  free massless fermions in an unbound space and
we introduce the notation used in this work.

Thermal states are constructed starting from the solution of the Dirac
equation with an external constant homogeneous magnetic field along
the $\hat{\vec{z}}$ direction.
Using cylindrical coordinates $\{t,\,r,\,\phi,\,z\}$ and the Dirac
representation of gamma matrices, the solution of the Dirac
equation in an external gauge field $A^\mu =(0, -B y/2, B x/2,0)$, such
that $q B>0$  with $q$ the charge of the fermion, is, in an unbound space \cite{book:SokolovAndTernov},
\begin{equation}
\label{eq:SolutionDirac}
\begin{split}
\psi_{\infty}=&\E^{-\I E_{n,p_z} t}\frac{\E^{\I p_z z}}{\sqrt{2\pi}}\frac{\E^{\I m \phi}}{\sqrt{2\pi}} \sqrt{|qB|} \\
&\times \left(\begin{array}{c}
	C_1 I_{n,a}(\rho) \E^{-\I\phi/2}\\
 \I C_2 I_{n-1,a}(\rho)\E^{+\I\phi/2} \\
	C_3 I_{n,a}(\rho)\E^{-\I\phi/2} \\
 \I C_4 I_{n-1,a}(\rho)\E^{\I\phi/2}
\end{array}\right),
\end{split}
\end{equation}
where $n=0,\,1,\,2,\,3,\dots$ is the principal quantum number,
the magnetic quantum number $m$ is the eigenvalue of the $z$-component of the total angular momentum,
$a=n+m-1/2=0,\,1,\,2,\,3,\dots$, and we defined the Laguerre functions
\begin{equation}
\label{eq:DefIfunc}
I_{n,a}(\rho)=\sqrt{\frac{a!}{n!}} \E^{-\rho/2} \rho^{\tfrac{n-a}{2}} L_a^{n-a}(\rho),
\end{equation}
with $\rho=|qB|r^2/2$ and $L_a^{n-a}(\rho)$ the Laguerre polynomials.
The state (\ref{eq:SolutionDirac}) is an eigenvector of the Hamiltonian with energy
\begin{equation*}
    E_{n,p_z} = \sqrt{2n |qB| + p_z^2 }.
\end{equation*}
The coefficients $C_1,\dots,\,C_4$ can be fixed by requiring that the state (\ref{eq:SolutionDirac})
is also an eigenvector of a chosen spin operator and that it is normalized according to
\begin{equation}
\label{eq:NormUnbounded}
\int \D^3 x\, \bar{\psi}_\infty \gamma^0 \psi_\infty =\sum_{i=1}^4 C_i^2=1.
\end{equation}
As the spin operator we choose the $z$-component of the magnetic spin $\h{\mu}_z$   defined as \cite{book:SokolovAndTernov}
\begin{equation}
\h{\vec{\mu}}= \vec{\Sigma}
    - \frac{\I\gamma_0 \gamma_5}{2}\vec{\Sigma}\times (\vec{p}-q\vec{A})\,.
\end{equation}
It is convenient for imposing the boundary conditions, as we show in the next section. Its eigenstates are referred to as the transverse polarizations.
Solving the secular equation
\begin{equation}
\label{eq:SecularTransverse}
\h{\mu}_z \,\psi_\infty = \zeta\, \sqrt{(E - \Omega \, m)^2 -p_z^2}\; \psi_\infty,
\end{equation}
with $\zeta=\pm$, and imposing  (\ref{eq:NormUnbounded}), we obtain
\begin{equation}
\label{eq:TransversePolUnbounded}
C_{1,3} = -\frac{1}{2\sqrt{2}}\left(A^+ \pm \zeta A^- \right),\quad
C_{2,4} = \frac{1}{2\sqrt{2}}\left(A^- \mp \zeta A^+ \right),
\end{equation}
with the upper signs referring to the indexes $1,\,2$ and the lower ones to $3,\,4$, and
\begin{equation*}
    A^\pm = \sqrt{1 \pm \frac{p_z}{E}} .
\end{equation*}
To compute the CME it is convenient to split the state into the
right- and left-handed parts
\begin{equation}
\psi_{\chi,\infty} = \frac{1+\chi\,\gamma^5}{2} \psi_{\infty},\quad
\chi=\pm=\text{R/L},
\end{equation}
where $\chi$ denotes the chirality of the state.
The chiral state is of the same form as in Eq. (\ref{eq:SolutionDirac}) but replacing the coefficients
$C_1,\dots,\,C_4$ with
\begin{equation*}
\begin{split}
C^\chi_{1,3} =& -\frac{1}{2\sqrt{2}}\left(\frac{1+\chi}{2} A^+ \pm \zeta \frac{1-\chi}{2} A^- \right),\\
C^\chi_{2,4} =& \frac{1}{2\sqrt{2}}\left(\frac{1+\chi}{2} A^- \mp \zeta \frac{1-\chi}{2} A^+ \right).
\end{split}
\end{equation*}

Considering only the particle part, the Dirac field with a definite chirality is
\begin{equation}
\h{\psi}_{\chi,\infty}(x)=\SumInt_{\hat{p}} \psi_{\chi,\infty}(x)\, \h{a}_\chi(\hat{p}),
\end{equation}
where $\hat{p}=\{n,\,m,\,p_z\}$. The thermodynamic equilibrium with a chiral imbalance
$\mu_A=\mu_R-\mu_L$ is described with the density operator
\begin{equation}
\h{\rho} = \frac{1}{Z}\E^{-\left(\beta \h{H} - \beta\mu_R \h{Q}_R -\beta\mu_L\h{Q}_L \right)}.
\end{equation}
Since we are considering a free-field, it readily follows that
\begin{equation}
\mean{\h{a}_{\chi'}^\dagger(\hat{p}')\h{a}_\chi(\hat{p})}
    = \delta_{\chi',\chi}\delta_{\hat{p}',\hat{p}}\, n_F (E_{\hat{p}}-\mu_\chi)
\end{equation}
where $n_F$ is the Fermi-Dirac distribution function with temperature $T=1/\beta$
\begin{equation}
    n_F(x) = \frac{1}{\E^{\beta x}+1} .
\end{equation}
The CME is obtained evaluating the thermal expectation value of the electric current.

We first compute the thermal expectation value of the right and left currents.
The electric current is then obtained from the right and left currents as a simple sum:
$\h{j}=\h{j}_R+\h{j}_L$.
The thermal expectation value of the particle contribution of the right current is
\begin{equation}
\label{eq:j3particle}
\begin{split}
\mean{\h{j}_R^\mu(\vec{x})}_+
=& \tr\left[ \h{\rho} \, \h{j}_{R\,+}^\mu(\vec{x})\right]\\
    =& \SumInt_{\hat{p}} n_F (E_{\hat{p}}-\mu_R) j^\mu_R(\vec{x}|\hat{p})\\
    =&\sum_{n,a=0}^\infty \int_{-\infty}^\infty \D p_z \,
         n_F (E_{\hat{p}}-\mu_R) j_R^\mu(\vec{x}|\hat{p}),
\end{split}
\end{equation}
where the matrix element is obtained with
\begin{equation}
j_R^\mu(\vec{x}|\hat{p}) = q\, \bar{\psi}_{R,\infty}(\vec{x}) \gamma^\mu \psi_{R,\infty}(\vec{x}).
\end{equation}
The antiparticle contribution can be added at the end of the calculation by changing
the sign of the chemical potential in the particle contribution.
Using the solution of the Dirac equation above, the component along the magnetic field is
\begin{equation}
j_R^z (\vec{x}|n,\,a,\,p_z) = \frac{q^2 B}{2\pi^2}\left[
    C_1^R C_3^R I^2_{n,a}(\rho) - C_2^R C_4^R I^2_{n-1,a}(\rho) \right]    
\end{equation}
with
\begin{equation}
C_{1,3}^R = \frac{1}{2\sqrt{2}}\sqrt{1 + \frac{p_z}{E_{n,p_z}}},\quad
C_{2,4}^R = \frac{1}{2\sqrt{2}}\sqrt{1 - \frac{p_z}{E_{n,p_z}}}.
\end{equation}
Considering that the odd terms in $p_z$ are vanishing once integrated, and there is no $\zeta$ dependence, we have
\begin{equation*}
\begin{split}
\mean{\h{j}_R^3(\vec{x})}_+ =& \frac{q^2 B}{8\pi^2} \sum_{n,a=0}^\infty
    \int_{-\infty}^\infty \D p_z \, n_F (E_{n,p_z}-\mu_R)\\
    &\times \left[I^2_{n,a}(\rho) - I^2_{n-1,a}(\rho)\right].
\end{split}
\end{equation*}
Including the antiparticles we obtain
\begin{equation*}
\begin{split}
\mean{\h{j}_R^3(\vec{x})} =& \frac{q^2 B}{4\pi^2} \sum_{n,a=0}^\infty
    \left[I^2_{n,a}(\rho) - I^2_{n-1,a}(\rho)\right]\\
    \times&\int_{0}^\infty \D p_z \left[ n_F (E_{n,p_z}-\mu_R) - n_F (E_{n,p_z}+\mu_R)\right].
\end{split}
\end{equation*}
The left current is obtained in the same fashion, and the electric current is the
sum of the two currents
\begin{equation}
\label{eq:CMESums}
\begin{split}
\mean{\h{j}^3(\vec{x})} =& \frac{q^2 B}{4\pi^2} \sum_{n,a=0}^\infty
    \left[I^2_{n,a}(\rho) - I^2_{n-1,a}(\rho)\right] \mathcal{I}_n(T,\mu,\mu_A),
\end{split}
\end{equation}
where
\begin{equation}
\begin{split}
\mathcal{I}_n =& \int_{0}^\infty\!\! \D p_z
    \left[ n_F (E_{n,p_z}-\mu_R) - n_F (E_{n,p_z}+\mu_R)\right.\\
    &\left.-  n_F (E_{n,p_z}-\mu_L) + n_F (E_{n,p_z}+\mu_L)\right].
\end{split}
\end{equation}
In view of the exponential falloff  of  the Fermi-Dirac distribution one can prove that $\mathcal{I}_n$ is finite and that the sum over $n$ converges absolutely. Therefore, we can first sum over $a$ and then over $n$.
Using the following properties of the $I$ functions (see, for instance, \cite{book:SokolovAndTernov}):
\begin{equation*}
I_{-1,a}(\rho)=0,\quad
\sum_a I^2_{n,a}(\rho)=1,
\end{equation*}
and noticing that $\mathcal{I}_n$ does not depend on $a$, it follows that
\begin{equation*}
\begin{split}
\mean{\h{j}^3(\vec{x})} =& \frac{q^2 B}{4\pi^2} \sum_{n,a=0}^\infty
    \left[I^2_{n,a}(\rho) - I^2_{n-1,a}(\rho)\right] \mathcal{I}_n\\
= & \frac{q^2 B}{4\pi^2} \sum_{a=0}^\infty \delta_{n,0}
    \left[I^2_{n,a}(\rho) - I^2_{n-1,a}(\rho)\right] \mathcal{I}_n \\
    & +\frac{q^2 B}{4\pi^2} \sum_{n=1}^\infty \sum_{a=0}^\infty
    \left[I^2_{n,a}(\rho) - I^2_{n-1,a}(\rho)\right] \mathcal{I}_n\\
= & \frac{q^2 B}{4\pi^2} \mathcal{I}_0 \sum_{a=0}^\infty I^2_{0,a}(\rho)\\
    & +\frac{q^2 B}{4\pi^2} \sum_{n=1}^\infty \mathcal{I}_n
    \left[\sum_{a=0}^\infty I^2_{n,a}(\rho) - \sum_{a=0}^\infty I^2_{n-1,a}(\rho)\right] \\
= & \frac{q^2 B}{4\pi^2} \mathcal{I}_0 +
    \frac{q^2 B}{4\pi^2} \sum_{n=1}^\infty \mathcal{I}_n
    \left[1 - 1\right] \\
= & \frac{q^2 B}{4\pi^2} \mathcal{I}_0 = \frac{q^2 B}{2\pi^2} \mu_A
\end{split}
\end{equation*}
where we used $\mathcal{I}_0=\mathcal{I}_0(T,\mu,\mu_A)=2\mu_A$.
This is how the CME (\ref{cme}) is derived from the exact solutions of the Dirac equation.
To include the finite-volume effects we follow the same procedure as described in this section but impose the boundary conditions.

\section{Solution of the Dirac equation in a finite cylinder}
\label{sec:DiracEq}
In this section, we solve the Dirac equation for a free massless fermion in a constant homogeneous magnetic field inside a cylinder of radius $R$ with the MIT boundary conditions.
The Dirac equation is the same as in the unbound case, but the requirements on the radial
part of the solutions are different. In the unbound case, one has to require that the wave
function is regular at the origin $\rho=0$ and vanishing at infinity $\rho\to\infty$.
In a cylinder of finite radius $R$ the wave function does not have to vanish at infinity.
Rather, we impose the boundary condition at $r=R$.
Solutions to the Dirac equation that are regular at the origin $r=0$
have the general form \cite{Chen:2017xrj,Sadooghi:2021upd}
\begin{equation}
\label{eq:SolutionDiracFiniteR}
\begin{split}
\psi=&\E^{-\I E_{n,p_z} t}\frac{\E^{\I p_z z}}{\sqrt{2\pi}}\frac{\E^{\I m \phi}}{\sqrt{2\pi}} \sqrt{|qB|}
\left(\begin{array}{c}
	C_1 F_1(\rho) \E^{-\I\phi/2}\\
 \I C_2 F_2(\rho)\E^{+\I\phi/2} \\
	C_3 F_1(\rho)\E^{-\I\phi/2} \\
 \I C_4 F_2(\rho)\E^{\I\phi/2}
\end{array}\right),
\end{split}
\end{equation}
where $F_{1,2}$ are given in terms of the confluent hypergeometric function:
\begin{align}
F_1(\rho) =& \frac{N_1}{\Gamma(m+\frac{1}{2})}\, \E^{-\rho/2} \rho^{\frac{m-\tfrac{1}{2}}{2}}
    \,_1F_1\left(-\lambda,\, m+\frac{1}{2};\, \rho \right), \nonumber\\
\label{eq:DefHyperF}
F_2(\rho) =&  \frac{N_2}{\Gamma(m+\frac{3}{2})}\, \E^{-\rho/2} \rho^{\frac{m+\tfrac{1}{2}}{2}}
    \,_1F_1\left(-\lambda+1,\, m+\frac{3}{2};\, \rho \right),
\end{align}
and $N_1$ and $N_2$ are such that
\begin{equation}
\begin{split}
\int_0^{\rho_R} \D\rho\,\left[F_1(\rho)\right]^2 =& 1,\quad
\int_0^{\rho_R} \D\rho\,\left[F_2(\rho)\right]^2 = 1.
\end{split}
\end{equation}

In an unbound case, the requirement that the wave function is vanishing at infinity
is fulfilled only if $n$ and $a=n+m-1/2$ are non-negative integers. In that case the
hypergeometric functions reduce to the Laguerre functions  (\ref{eq:DefIfunc}) and we recover the
unbound solutions (\ref{eq:SolutionDirac}). In a finite cylinder the principal quantum number $n$ does not have to be an integer. To stress this point, from now on, we denote the principal quantum number as $\lambda$. The magnetic quantum number $m$ is not affected by the boundary and runs over all semi-integer values. 
The energy in a finite cylinder is
\begin{equation}
\label{eq:EnergyFiniteCylinder}
    E_{\lambda,p_z} = \sqrt{2\lambda |qB| + p_z^2 },
\end{equation}
with the values of $\lambda$ to be determined according to the boundary condition.

At first glance it may seem that the simplest boundary condition is the Cauchy one that requires vanishing of the wave function (\ref{eq:SolutionDiracFiniteR}) on the surface $\rho=\rho_R=|qB|R^2/2$. However, it is easy to realize,
that for a finite $\rho_R$ the radial wave functions in Eq. (\ref{eq:DefHyperF}) cannot
both be set to zero for the same value of $\lambda$ and $m$. A more elaborate boundary
condition must be used. In this work we use the MIT boundary conditions which allow us 
to keep the system in a finite region without including an (explicit) confining potential
in the Hamiltonian. In this way the fields remain free.

The coefficients $C_1,\dots,\,C_4$ in the transverse polarization are obtained as described in the previous section:
\begin{equation}
\label{eq:CoefficientsTransversePol1Bounded}
\begin{split}
C_{1,3}=&\frac{1}{2}\frac{N_1 N_2}{\sqrt{\lambda N_1^2 + N_2^2}} \frac{A_+ \pm \zeta A_-}{N_1},\\
C_{2,4}=&\frac{1}{2}\frac{N_1 N_2}{\sqrt{\lambda N_1^2 + N_2^2}}\frac{A_- \mp \zeta A_+}{N_2}\sqrt{\lambda},
\end{split}
\end{equation}
where $\zeta=\pm$ and
\begin{equation}
\label{eq:CoefficientsTransversePol2Bounded}
A_\pm =  \sqrt{1 \pm\frac{p_z}{E_{\lambda,p_z}} }.
\end{equation}
The exact same coefficients are obtained if we replace the condition
in Eq. (\ref{eq:SecularTransverse}) with the requirement that
\begin{equation}\label{eq:cccc}
\frac{C_4}{C_1}=-\frac{C_3}{C_2}.
\end{equation}
As we show below, this is a necessary condition for a state to satisfy the MIT boundary condition.
Therefore, the MIT boundary condition can be applied directly to states in the transverse polarization.
Different spin states can be chosen, such as helicity states, but in that case, only a specific linear
combination of spin states can satisfy the MIT boundary condition~\cite{Ambrus:2015lfr}. 

\subsection{The MIT boundary condition}
\label{sec:MITBc}
The MIT bag model \cite{Chodos:1974je,Shanker:1983yd} allows us to find solutions
of the Dirac equation in a bound region without introducing discontinuities in the
wave function and preserving the conserved quantities of the system, such as energy,
momentum and electric charge, thanks to the vanishing of the fluxes of the corresponding
currents at the boundary. Ultimately, all these features are consequences of the
self-adjointness of the Hamiltonian with respect to the Dirac inner product \cite{Ambrus:2015lfr,Sitenko:2016iqb}.

Given any two solutions of the Dirac equation $(\psi,\,\Phi)$, the Dirac inner product
is defined as
\begin{equation}
\langle \psi,\,\Phi\rangle = \int_{V} \D^3x\,\bar{\psi}(x) \gamma^0 \Phi(x).
\end{equation}
The Hamiltonian is a self-adjoint operator if
\begin{equation}
\langle \psi,\,\h{H}\Phi\rangle = \langle \h{H}\psi,\,\Phi\rangle,
\end{equation}
or, equivalently, for $\h{H}=\I\de_t$, when the inner product is time independent
\begin{equation}
\de_t \langle \psi,\,\Phi\rangle = 0.
\end{equation}
Integrating by parts, this condition can be written as an integral over the boundary of
the volume
\begin{equation}
-\int_{\de V}\D\Sigma_i  \,\bar{\psi}(x) \gamma^i \Phi(x) = 0,
\end{equation}
which in the case of a cylinder of radius $R$ reads
\begin{equation}
\label{eq:Hselfadj}
\begin{split}
R\int_0^\infty\!\!\!\D z\int_0^{2\pi}\!\!\!\D\phi
    \,\bar{\psi}(R) \gamma^r \Phi(R) =& 0,\\
\de_r (\bar\psi\Phi)(R)=&- \Theta(R),
\end{split}
\end{equation}
where  $\psi(R)=\psi(z,\,\phi,\, r=R)$, $\Theta(R)<\infty$
and $\gamma^r =\slashed{n}=\hat{n}_\mu \gamma^\mu =\hat{r}_\mu \gamma^\mu$, with $\hat{n}=\hat{r}$
the radial direction in cylindrical coordinates.

In order to have a self-adjoint Hamiltonian we require that any solution $\psi$
of the Dirac equation is also solving the MIT boundary condition
\begin{equation}
\label{eq:MITBC}
\I \gamma^r \psi(R) = - \psi(R),
\end{equation}
Indeed, this condition and its conjugate $\I \bar\psi(R) \gamma^r = \bar\psi(R)$
imply \cite{Chodos:1974je,Shanker:1983yd}
\begin{equation}
\begin{split}
\bar\psi(R)\Phi(R)=& \bar\psi(R)[-\I \slashed{n}\Phi(R)]\\
    =&-[\I \bar\psi(R)\slashed{n}]\Phi(R)
    = -\bar\psi(R)\Phi(R) = 0,
\end{split}
\end{equation}
from which it follows that the integrand of Eq. (\ref{eq:Hselfadj}) is vanishing at any point $z,\,\phi$:
\begin{equation}
\bar{\psi}(R) \slashed{n} \Phi(R) = -\I \bar{\psi}(R) \Phi(R) = 0.
\end{equation}
Using the MIT boundary conditions is not the only way one can require the vanishing
of Eq.~(\ref{eq:Hselfadj}). In \cite{Chen:2017xrj,Sadooghi:2021upd} this is
ensured requiring that
\begin{equation}
\label{eq:GlobalBC}
F_1(\lambda,\, m,\, \rho_R) F_2(\lambda',\, m,\, \rho_R) = 0
\end{equation}
for arbitrary values of $m$, $\lambda$, and $\lambda'$, which is
the generalization to the case of the magnetic field of the spectral
boundary conditions discussed in \cite{Ambrus:2015lfr}.

From the self-adjointness of the Hamiltonian we can also show that the normalized eigenvectors
of the Hamiltonian are orthogonal with respect to the Dirac inner product.
Take two eignevectors of the Hamiltonian $\Phi$ and $\psi$ with quantum numbers
$\hat{p}=\{\lambda,\, m,\, p_z,\, \zeta\}$ and $\hat{p}'=\{\lambda',\, m',\, p'_z,\, \zeta'\}$
respectively. Using
\begin{equation}
\begin{split}
\de_t \langle \psi(t),\Phi(t) \rangle
    =& \de_t  \left[\E^{-\I(E_{\lambda}-E_{\lambda'})}\langle \psi(t=0),\Phi(t=0) \rangle\right]\\
    =&-\I(E_{\lambda}-E_{\lambda'})\langle \psi(t),\Phi(t) \rangle=0,
\end{split}
\end{equation}
it follows that when $E_{\lambda}\neq E_{\lambda'}$ we must have $\langle \psi,\Phi \rangle=0$.
The integrals over $z$ and $\phi$ immediately give
\begin{equation}
\begin{split}
\langle \psi,\Phi \rangle =& \delta(p_z-p_z')\delta_{m,m'} \int_0^{\rho_R}\D\rho\,
        \E^{-\I(E_{\lambda}-E_{\lambda'})} \\
    &\times\left[H_1 F_1 F'_1 + H_2 F_2 F'_2\right]
\end{split}
\end{equation}
where
\begin{equation}
\begin{split}
H_1 =& \frac{N_2 N'_2}{\sqrt{\lambda N_1^2 + N_2^2}\sqrt{\lambda' N_1^{\prime 2} + N_2^{\prime 2}}}
    \frac{A_+ A'_+ + \zeta\zeta' A_- A'_-}{2},\\
H_2 =& \frac{\sqrt{\lambda\lambda'}N_1 N'_1}{\sqrt{\lambda N_1^2 + N_2^2}\sqrt{\lambda' N_1^{\prime 2} + N_2^{\prime 2}}}
    \frac{A_- A'_- + \zeta\zeta' A_+ A'_+}{2}.
\end{split}
\end{equation}
From what we showed above the remaining integral must give $\delta_{j,j'}\delta_{\zeta,\zeta'}$.
Because the wave functions have been normalized, we have
\begin{equation}
\langle \psi,\Phi \rangle =  \int_{V} \D^3x \, \bar{\psi}_{\hat{p}'}\gamma^0 \Phi_{\hat{p}}
= \delta(p_z-p_z')\delta_{m,m'} \delta_{\zeta,\zeta'}\delta_{j,j'}.
\end{equation}

We already found that the general form of the solution of the Dirac equation is
given by (\ref{eq:SolutionDiracFiniteR}). Imposing the condition (\ref{eq:MITBC}) to the
general solution of the Dirac equation (\ref{eq:SolutionDiracFiniteR}), we obtain
\begin{equation}
\begin{cases}
C_1 F_1(\rho_R) - C_4 F_2(\rho_R) = 0, \\
C_3 F_1(\rho_R) + C_2 F_2(\rho_R) = 0,
\end{cases}
\end{equation}
whence
\begin{equation}
\label{eq:MITF1F2}
F_1(\rho_R) = \frac{C_4}{C_1} F_2(\rho_R)
	=  - \frac{C_2}{C_3} F_2(\rho_R),
\end{equation}
where $\rho_R$ is defined in (\ref{rhoR}).
The effect of the boundary on the energy spectrum increases as $\rho_R$ decreases. Since (\ref{eq:MITF1F2}) coincides with (\ref{eq:cccc}), our choice of the transverse polarization basis is the most convenient, as mentioned above.

If $F_1$ or $F_2$ vanish identically, we cannot impose the MIT boundary condition. In particular, this is the case for $\lambda=0$ because then $F_2=0$.  In the unbound case the CME is caused entirely by the lowest Landau Level (LLL)
$n=0$ which is also the chiral state. With the MIT boundary condition the LLL is not realized
for $\lambda=0$, and all eigenstates are not chiral. Therefore, at this point of the analysis, it
is not clear whether the CME can also be realized in a finite region with this type of boundary condition.
Solutions with $\lambda$ very close to zero but still positive are instead allowed, and the new
ground state replacing the lowest Landau level is discussed below.

As we already indicated, the boundary condition can only be satisfied by states such that $C_2/C_3=-C_4/C_1$,
which is always the case for the transverse polarization. Therefore, using states in the transverse
polarization with coefficients (\ref{eq:CoefficientsTransversePol1Bounded}),
the MIT boundary condition becomes
\begin{equation}
\label{eq:MITConstraint}
\frac{F_1(\rho_R)}{N_1} = \zeta \sqrt{\lambda}\frac{F_2(\rho_R)}{N_2}.
\end{equation}
Solving this equation for $\lambda$ gives the energy spectrum (\ref{eq:EnergyFiniteCylinder})
and the whole set of normal modes (\ref{eq:SolutionDiracFiniteR}). Equation (\ref{eq:MITConstraint})
has an infinite countable number of roots $\{\lambda_i\}$, $i=1,2,3,\dots$, where
$\lambda_i < \lambda_{i+1}$. These roots generally depend on $m,\,\zeta$, and $\rho_R$.
As $\rho_R$ approaches infinity the roots approach the values
$\lambda=n=0,\,1,\,2,\dots$ independently of $m$ and $\zeta$.

It is instructive to find the analytic value of the principal quantum number $\lambda$
corresponding to the LLL as the radius of the cylinder is increased.
As said, when increasing the parameter $\rho_R$, $\lambda$ becomes an integer and one can check that
the functions $F_1$ and $F_2$ vanish for $a< 0$. For $\lambda\ll  1$, this implies that $m\geq +1/2$.
In this case, we can expand the constraint (\ref{eq:MITConstraint}) in the power series of $\lambda$
and obtain the equation
\begin{equation}
\sqrt{\lambda} = \zeta \frac{\E^{-\rho_R}\rho_R^{m}}{\Gamma\left(m+\tfrac{1}{2}\right)-\Gamma\left(m+\tfrac{1}{2},\rho_R\right)}.
\end{equation}
It has a solution only when $\zeta=+1$:
\begin{equation}
\label{eq:LambdaFirstRoot}
\lambda_{\rm LLL} \approx \frac{\E^{-2\rho_R}\rho_R^{2m}}{\Gamma\left(m+\tfrac{1}{2}\right)^2}.
\end{equation}
This solution is finite, even though it approaches zero as the radius is increased.
This is different from the unbound case where the LLL $\lambda=0$ exists for both polarizations $\zeta=\pm$. Nevertheless, the unbound CME is recovered in the large radius limit. This occurs because the energy levels are weighted more in the bound case
compared to the unbound one.
Indeed, we can compute analytically the normalization constants $N_1$ and $N_2$ for $\lambda=0$ and derive that in the infinite cylinder limit, the bound modes become
\begin{equation*}
\lim_{\rho_R\to \infty} \psi_{a>0,\zeta} = \begin{cases}
     \psi_{\zeta,\infty} & \lambda \geq 1, \\
     0 & \lambda=0,\quad \zeta=-1, \\
     \sqrt{2}\psi_{\lambda=0,a,\zeta=+1,\infty} & \lambda=0,\quad \zeta=+1 ,
\end{cases}
\end{equation*}
where $\psi_{\zeta,\infty}$ are the unbound solutions given in Eq. (\ref{eq:SolutionDirac}).

\begin{figure*}[ht]
\includegraphics[width=0.45\textwidth]{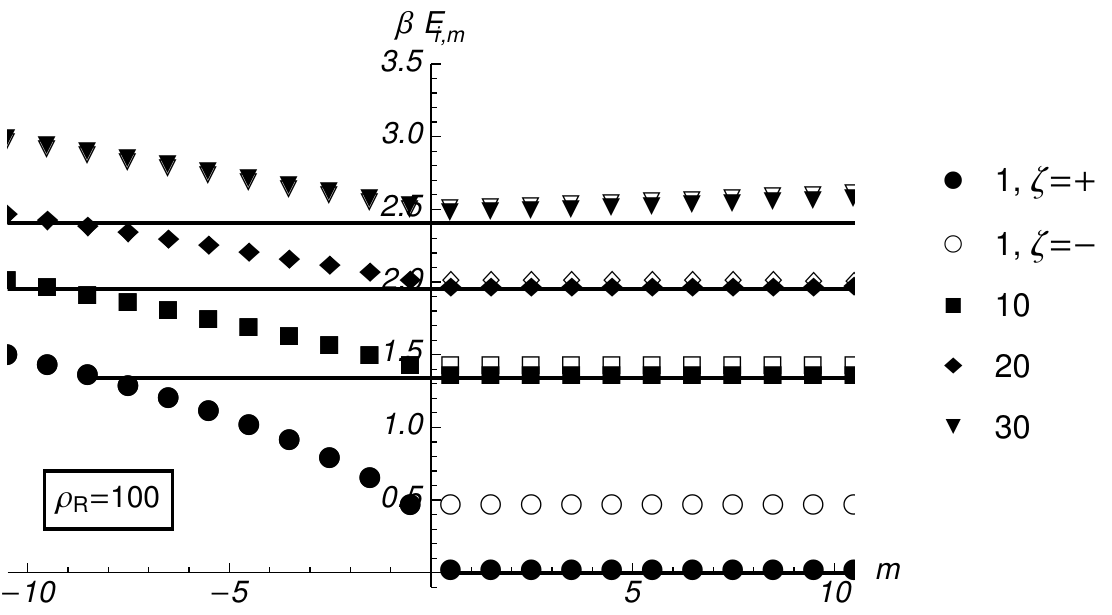}
\includegraphics[width=0.45\textwidth]{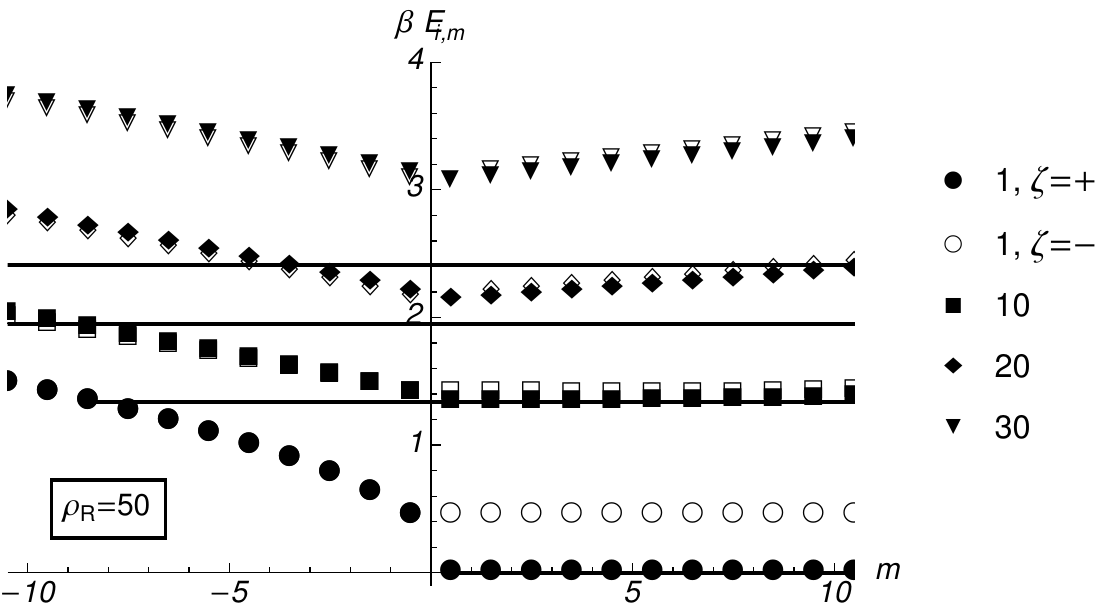}\\
\includegraphics[width=0.45\textwidth]{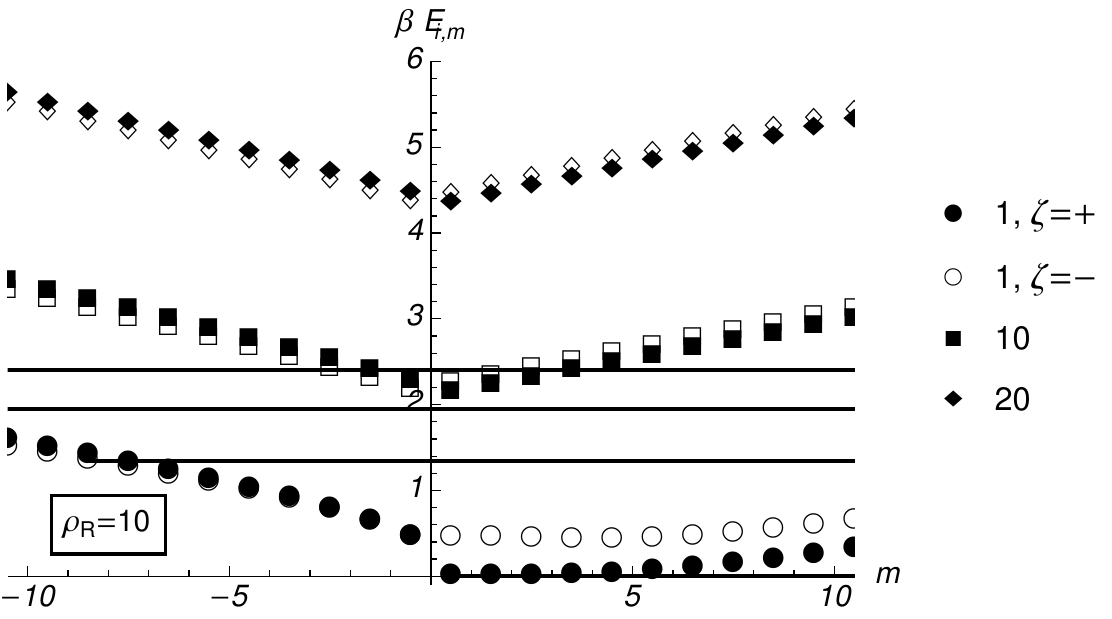}
\includegraphics[width=0.45\textwidth]{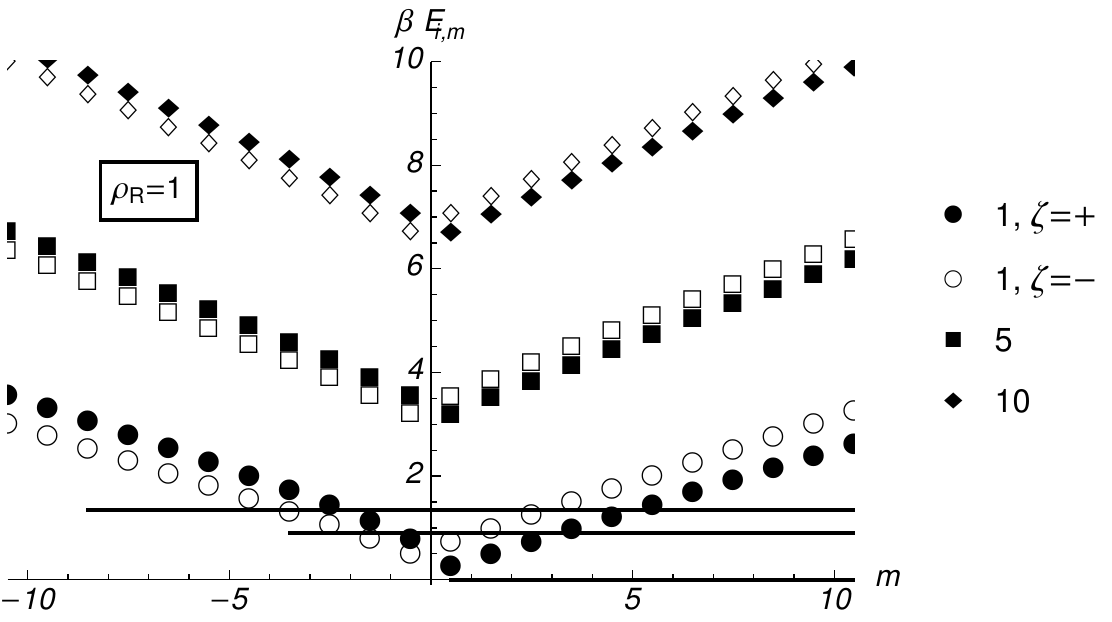}\\
\includegraphics[width=0.45\textwidth]{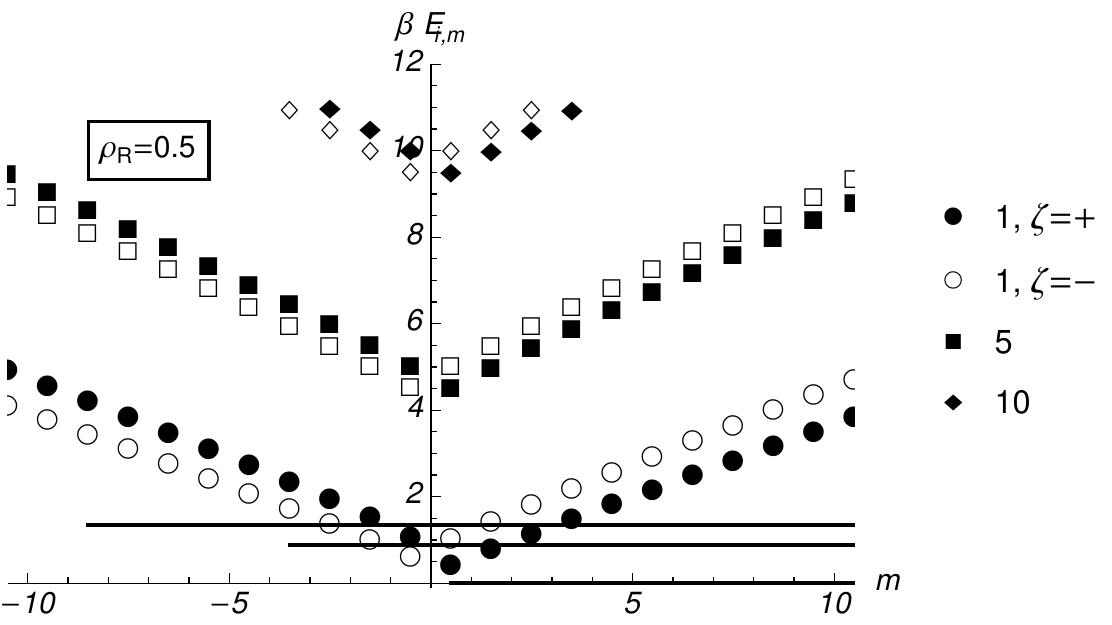}
\includegraphics[width=0.45\textwidth]{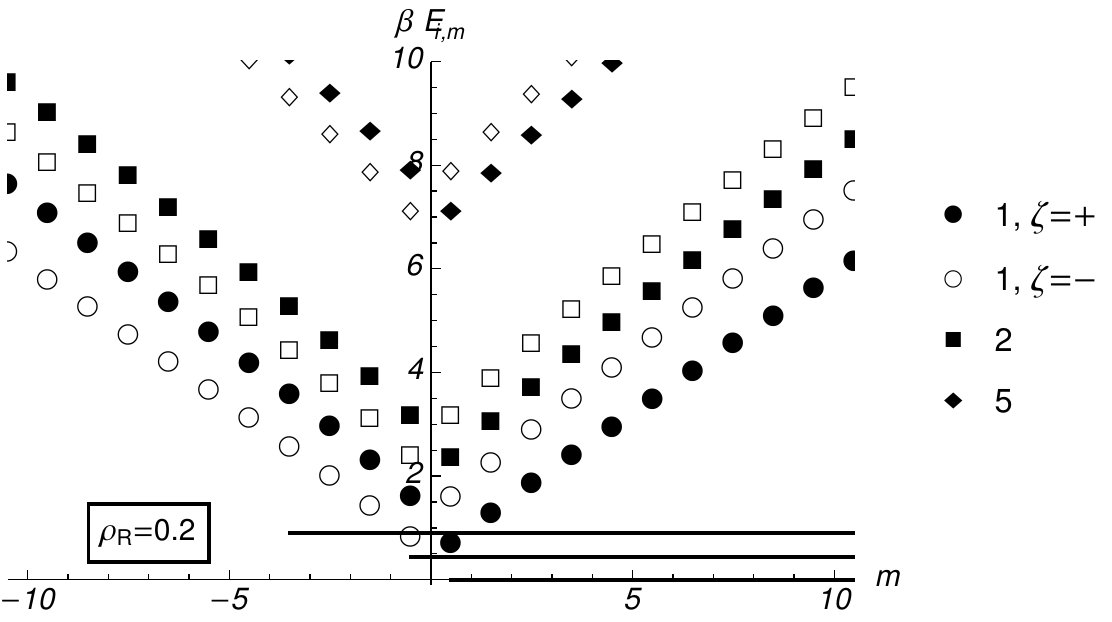}
\caption{The energy levels (\ref{eq:EnergyLevel}) as a function of $m$ for different values
of $\rho_R$ at $\beta^2|qB|=0.1$, and $p_z=0$. Different markers correspond
to the $i$'th root of Eq. (\ref{eq:MITConstraint}) as indicated in the legends. 
Solid markers correspond to positive polarization $\zeta=+1$, and open markers to $\zeta=-1$.
Solid lines are the energy levels of the unbound case, $\beta E^\infty_{i,m}=\theta(a)\sqrt{2(i-1)\beta^2|qB|}$, corresponding to the Landau levels plotted in the same panel. The ordinate range varies for different panels.}
\label{fig:EnergySpectrum}
\end{figure*}

To obtain the energy spectrum, we numerically compute the roots $\{\lambda_k\}$ of (\ref{eq:MITConstraint})
at  fixed $\rho_R$, $\zeta$, and $m$.
Figure~\ref{fig:EnergySpectrum} shows the energy levels
\begin{equation}
\label{eq:EnergyLevel}
\beta E_{n,i}=\sqrt{2\lambda_i \beta^2 |qB|},    
\end{equation}
with $p_z=0$ and $\beta^2|qB|=0.1$,
where $\beta$ is a length scale, for instance the inverse temperature, and  $i$ denotes the $i$-th root of $\lambda$. These results are in
agreement with those of \cite{Chen:2017xrj,Sadooghi:2021upd}; see also \cite{Mameda:Thesis} for a
detailed discussion, where solutions of
free fermions in a magnetic field in a finite cylinder were presented with the boundary condition (\ref{eq:GlobalBC}).

\subsection{Thermodynamics in a finite cylinder}
\label{sec:Thermo}

In this section, we show that the energy, momentum, electric and axial charge are conserved
inside the cylinder when considering solutions of the Dirac equation with the MIT boundary condition.
This ensures that the equilibrium density matrix describes the thermal ensemble of the system and
we can use statistical mechanics to evaluate the chiral magnetic effect as done in the 
case of the unbound region.

The equilibrium thermal state of a system is described by the equilibrium statistical operator
\begin{equation}
\label{eq:StatOper}
\h{\rho} = \frac{1}{Z} \exp\left\{- \h{H}/T +\mu\, \h{Q}/T +\mu_A \h{Q}_A/T  \right\},
\end{equation}
provided that the following quantities are conserved:
\begin{equation}
\begin{split}
\h{P}^\nu=&\int_V \D^3 x\, \h{T}^{0\nu},\quad
\h{Q}= \int_V \D^3 x\, \h{j}^0,\\
\h{Q}_A=& \int_V \D^3 x\, \h{j}_A^0,
\end{split}
\end{equation}
which are, respectively, the total four-momentum, the electric charge and the axial charge.
Even if the region is finite, these quantities are conserved if the fluxes at the boundary vanish
\begin{equation}
\begin{split}
\int_{\de V} \D S\, \h{T}^{i\nu} \hat{n}_i=& 0,\quad
\int_{\de V} \D S\, \h{j}^i \hat{n}_i =0,\\
\int_{\de V} \D S\, \h{j}_A^i \hat{n}_i=& 0. 
\end{split}
\end{equation}
In the case of a cylinder, $\hat{n}$ is the radial direction.

We now explicitly show that the MIT boundary condition ensures that all the fluxes are vanishing,
not just the electric current. We first remind the reader that the MIT boundary condition implies
\begin{equation}
\I \slashed{n} \psi(R) = - \psi (R),\quad
\bar\psi\psi(R)=0,\quad \de_r (\bar\psi\psi)(R)=- \Theta(R).
\end{equation}
From the general form of the solutions (\ref{eq:SolutionDiracFiniteR}), we also have that
\begin{equation}
\bar\psi\gamma^5\psi=0.
\end{equation}
Using these relations it is easy to prove that the electric current is conserved:
\begin{equation}
\hat{n}_\mu \h{j}^\mu (R) = \bar\psi(R)\slashed{\hat{n}} \psi(R) = \I \bar\psi\psi(R)=0.
\end{equation}
Similarly, for the axial current we obtain
\begin{equation}
\begin{split}
\hat{n}_\mu \h{j}_A^\mu (R) =& \bar\psi(R)\slashed{n}\gamma^5 \psi(R)
    = -\bar\psi(R)\gamma^5\slashed{\hat{n}} \psi(R) \\
    =& -\I \bar\psi\gamma^5\psi(R)=0.
\end{split}
\end{equation}
Finally, we also check that the flux of energy and momentum is vanishing.
Using the canonical energy-momentum tensor of a Dirac field interacting with an external
gauge field we have:
\begin{equation}
\begin{split}
\int_{\de V} & \!\!\!\D S\, \h{T}^{\mu\nu} \hat{n}_\mu = \int_{\de V} \!\!\!\D S \left[\frac{\I}{2}\left(
    \bar\psi\slashed{\hat{n}}\de^\nu\psi - \de^\nu\bar\psi\slashed{\hat{n}}\psi\right)
    - \bar\psi\slashed{\hat{n}}\psi A^\nu\right]\\
= & \int_{\de V}\D S\, \left[\frac{1}{2}\left(\bar\psi\de^\nu\psi + \de^\nu\bar\psi\psi \right)
    -\I\bar\psi\psi(R)A^\nu\right]\\
= & \frac{1}{2}\int_{\de V}\D S\,\de^\nu(\bar\psi\psi)
    = \frac{1}{2} \Theta(R) \int_{\de V}\D S\, \hat{n}^\nu =0,
\end{split}
\end{equation}
where we used $\bar\psi\psi(R)=0$.
\newpage

\section{The chiral magnetic effect in a finite cylinder}
\label{sec:CME}
The CME is obtained by evaluating the thermal average of the electric current with the statistical
operator (\ref{eq:StatOper}), as done in Sec. \ref{sec:Review}.
This time we use the bound solutions (\ref{eq:SolutionDiracFiniteR}) in the
transverse polarization (\ref{eq:CoefficientsTransversePol1Bounded}), and satisfy the MIT boundary conditions (\ref{eq:MITConstraint}).
We first obtain the thermal expectation value of the right-handed current. The right-handed
projection of the solution of the Dirac equation with the MIT boundary condition is
\begin{widetext}
\begin{equation}
\label{eq:SolutionDiracBoundedTransverseR}
\psi_{R\zeta}=\E^{-\I\epsilon E t}\frac{\E^{\I p_z z}}{\sqrt{2\pi}}\frac{\E^{\I m \phi}}{\sqrt{2\pi}} \frac{\sqrt{|qB|}}{2}\frac{N_1 N_2}{\sqrt{\lambda N_1^2 + N_2^2}}
\left(\begin{array}{c}
	\sqrt{1+\frac{p_z}{E_{\lambda,p_z}}}\frac{F_1(\rho)}{N_1} \E^{-\I\phi/2}\\
 \I \sqrt{1-\frac{p_z}{E_{\lambda,p_z}}}\frac{\sqrt{\lambda}F_2(\rho)}{N_2}\E^{+\I\phi/2} \\
	\sqrt{1+\frac{p_z}{E_{\lambda,p_z}}}\frac{F_1(\rho)}{N_1} \E^{-\I\phi/2} \\
 \I \sqrt{1-\frac{p_z}{E_{\lambda,p_z}}}\frac{\sqrt{\lambda}F_2(\rho)}{N_2}\E^{\I\phi/2}
\end{array}\right).
\end{equation}
Therefore, the matrix element of the right-handed current is
\begin{equation}
\begin{split}
j^3_R (\vec{x}|\lambda,\,m,\,p_z,\,\zeta) =& q\, \bar{\psi}_{R\zeta}(\vec{x}) \gamma^\mu \psi_{R\zeta}(\vec{x})\\
=& \frac{q^2 B}{8\pi^2} \frac{\bar{N}^2}{2}\frac{\Gamma(1+a)}{\Gamma(1+\lambda)}
    \left[\frac{F_1^2(\rho)}{N_1^2} - \frac{\lambda F_2^2(\rho)}{N_2^2}
    +\frac{p_z}{E_{\lambda,p_z}}\left(\frac{F_1^2(\rho)}{N_1^2}
    + \frac{\lambda F_2^2(\rho)}{N_2^2}\right)\right],
\end{split}
\end{equation}
\end{widetext}
where
\begin{equation}
\bar{N}^2=\frac{ 2 N_1^2 N_2^2}{\lambda N_1^2 + N_2^2}\frac{\Gamma(1+\lambda)}{\Gamma(1+a)} 
\end{equation}
and $a=n+m-1/2$.
The thermal expectation value of the particle part of the right-handed current is
\begin{equation}
\begin{split}
\mean{\h{j}_R^3&(\vec{x})}_+ = \tr\left[ \h{\rho} \, \h{j}_{R\,+}^\mu(\vec{x})\right]
=\sum_{m=-\infty}^\infty \sum_{\zeta=\pm} \sum_{k=1}^\infty\\
&\times\int_{-\infty}^\infty \D p_z \, n_F (E_{\lambda_k,p_z}-\mu_R) j_R^3(\vec{x}|\lambda_k,\,m,\,p_z,\,\zeta).
\end{split}
\end{equation}
The summation over the roots $\{\lambda_k\}$ must be done for specific values of $m$ and $\zeta$
since the energy spectrum depends on these quantities.
Removing the $p_z$-odd part of the matrix element, which vanishes upon integration, and
including the antiparticle contribution, we obtain
\begin{equation}
\begin{split}
\mean{\h{j}_R^3(\vec{x})}=& \frac{q^2 B}{4\pi^2}
    \sum_{m,\zeta,k} \frac{\bar{N}^2}{2}\frac{\Gamma(1+a)}{\Gamma(1+\lambda_k)}\left[\frac{F_1^2(\rho)}{N_1^2} - \frac{\lambda_k F_2^2(\rho)}{N_2^2} \right]\\
&\times\!\int_0^\infty \!\!\!\D p_z  \left[ n_F (E_{\lambda_k,p_z}\!\!-\mu_R)\! - n_F (E_{\lambda_k,p_z}\!\!+\mu_R)\right].
\end{split}
\end{equation}
We compute the left-handed current in a similar way. The sum of the right- and left-handed currents yields the vector current:
\begin{equation}
\label{eq:CMEAnalytic}
\begin{split}
\mean{\h{j}_V^3(\vec{x})}=& \frac{q^2 B}{4\pi^2}
    \sum_{m,\zeta,k} \frac{\bar{N}^2}{2}\frac{\Gamma(1+a)}{\Gamma(1+\lambda_k)}\\
&\times\left[\frac{F_1^2(\rho)}{N_1^2} - \frac{\lambda_k F_2^2(\rho)}{N_2^2} \right]
    \mathcal{I}_{\lambda_k}(T,\mu,\mu_A) ,
\end{split}
\end{equation}
where we denoted
\begin{equation}
\label{eq:ThermalI}
\begin{split}
\mathcal{I}_\lambda(T,\mu,\mu_A) =& \int_{0}^\infty\!\! \D p_z
    \left[ n_F(E_{\lambda,p_z}\!-\!\mu_R) - n_F (E_{\lambda,p_z}\!+\!\mu_R)\right.\\
    &\left.-  n_F (E_{\lambda,p_z}\!-\mu_L) + n_F (E_{\lambda,p_z}\!+\mu_L)\right]\\
= & \mathcal{I}^R_\lambda(T,\mu_R) - \mathcal{I}^L_\lambda(T,\mu_L).
\end{split}
\end{equation}
In the infinite radius limit, Eq. (\ref{eq:CMEAnalytic}) reduces to 
(\ref{eq:CMESums}) because  $\lambda_k=n$, $\bar{N}^2=1$, and
\begin{equation}
\begin{split}
\frac{a!}{n!} \frac{F_1^2(\rho)}{N_1^2} =&
    I_{a,n}^2(\rho) = I_{n,a}^2(\rho), \\
\frac{a!}{n!} n \frac{F_2^2(\rho)}{N_1^2} = &
    I_{a,n-1}^2(\rho) = I_{n-1,a}^2(\rho).
\end{split}
\end{equation}

Equation~(\ref{eq:CMEAnalytic}) expresses the CME in a finite volume and is our main analytical result. 
The summation and the integration are converging and finite because the boundary condition
ensures that the energy spectrum is bound from below. In contrast  from the infinite-volume
case, where the CME is completely determined by the lowest Landau level, we observe in (\ref{eq:CMEAnalytic}) that in the finite volume the higher Landau levels ($k>1$)  also contribute to the CME.

Since the analytical roots of (\ref{eq:MITConstraint}) are unknown, the functional dependence of the bound CME (\ref{eq:CMEAnalytic}) on the model parameters can only be determined numerically. It is convenient to study the effect of the boundary by considering the
normalized value of the CME, which we define as the electric current in a finite region obtained from  (\ref{eq:CMEAnalytic}) divided by the CME conductivity in an unbound space:
\begin{equation}
\label{eq:NormCME}
\text{Normalized CME} = \frac{\mean{\h{j}_V^3(\vec{x})}}{q^2 B \mu_A /(2\pi^2)}.
\end{equation}
%

\subsection{Results}
\label{sec:CMEResults}
We computed the bound CME by truncating the sums in $m$ and $k$, and numerically evaluating  (\ref{eq:ThermalI}). We estimated the accuracy of the
calculation as the next term in the summation and we stopped the calculation when the error was smaller than
the thickness of the lines used to plot the results.
We found that the dependence on the axial chemical potential $\mu_A$ is linear as in the unbound chiral magnetic
effect. It therefore cancels out in the ratio (\ref{eq:NormCME}). With this observation in mind we set 
 $\beta\mu_A=0.1$ throughout the numerical analysis.

\begin{figure}[hbt]
\centering
\includegraphics[width=0.45\textwidth]{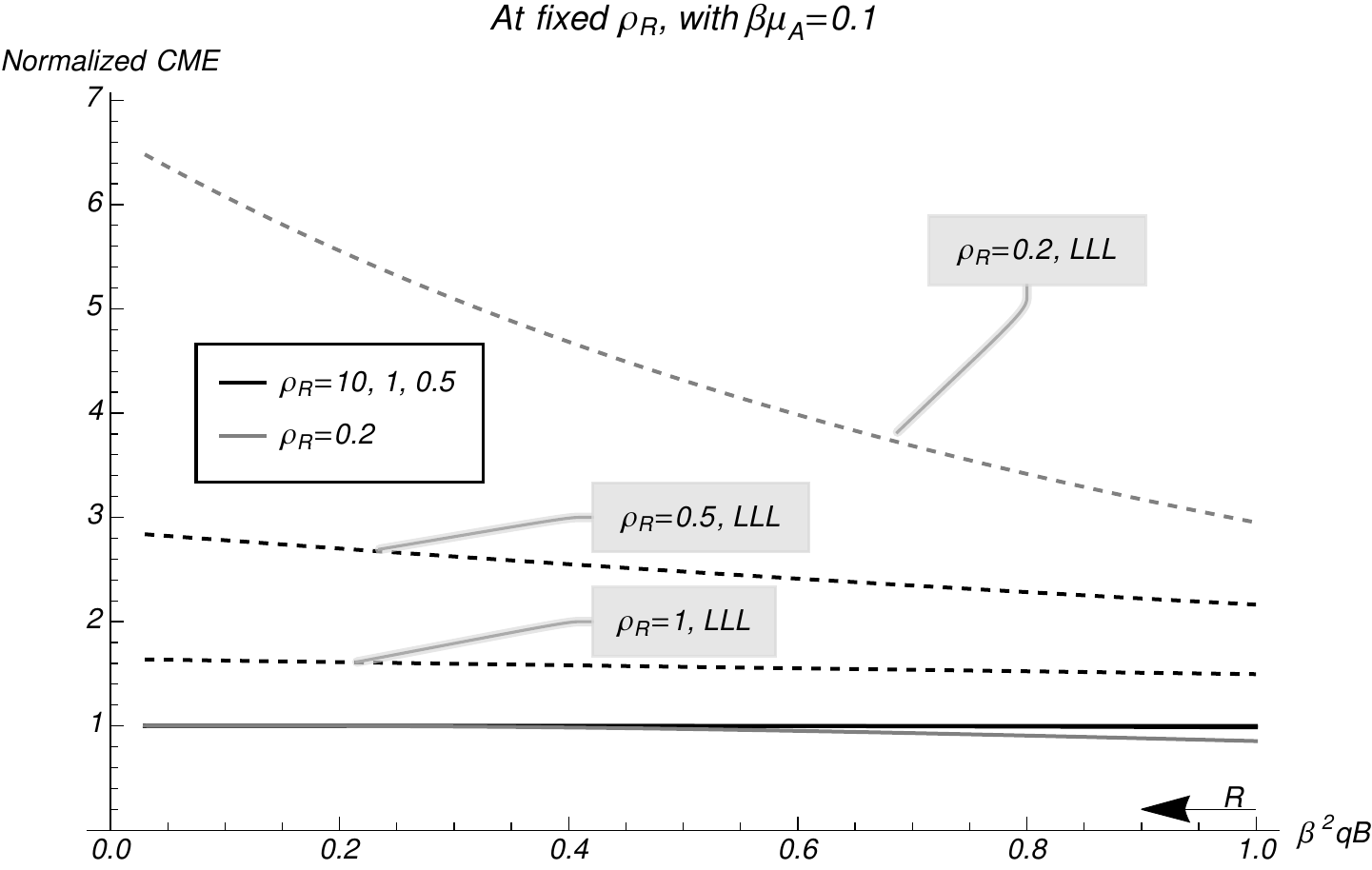}\\
\includegraphics[width=0.45\textwidth]{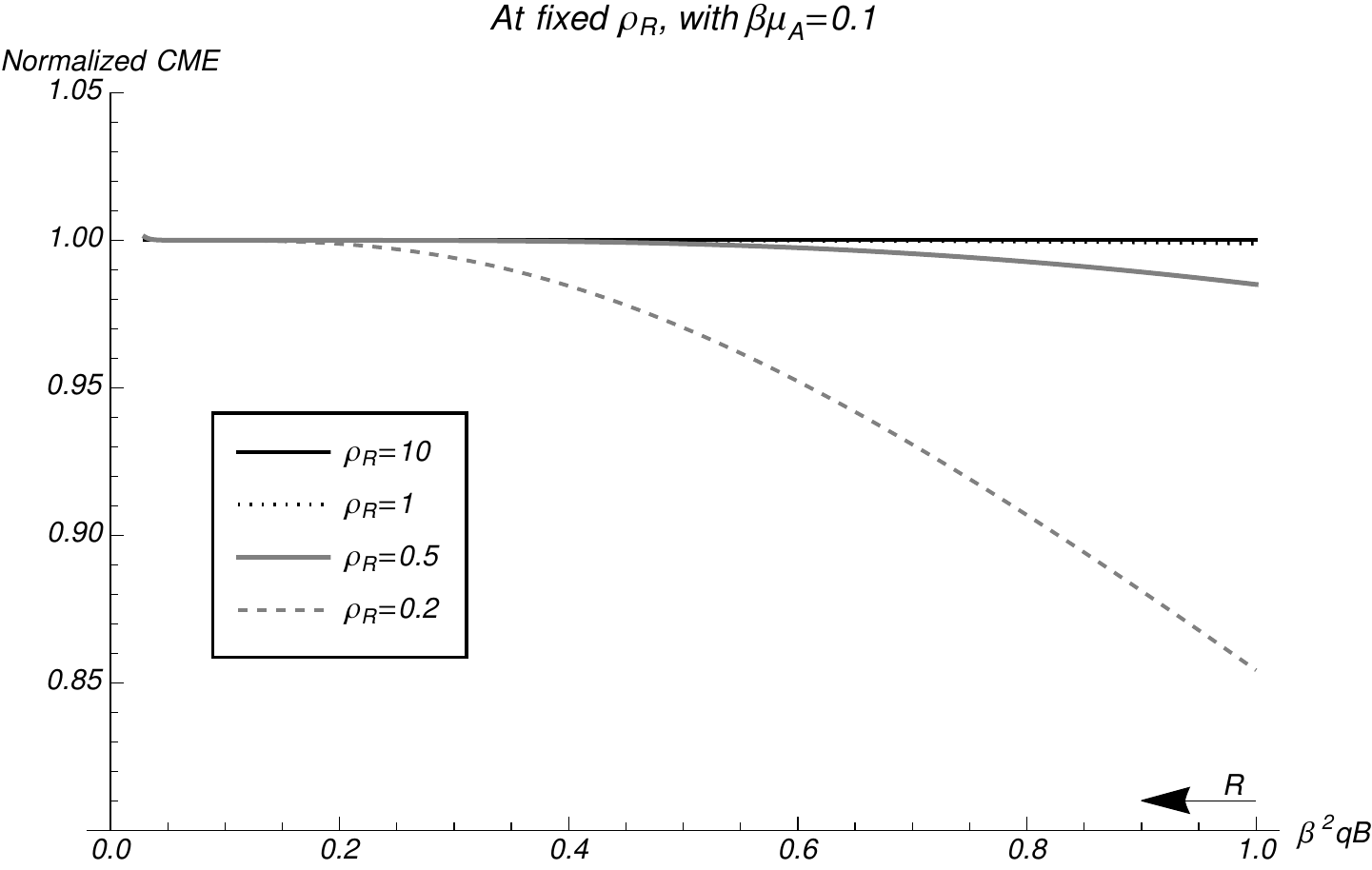}
\caption{The normalized CME (\ref{eq:NormCME}) at fixed $\rho_R$ and different values of the magnetic field $|qB|$/radius $R$. Solid lines are the total contributions, while dashed lines are the contribution of the LLL. The bottom panel is the same plot as the one in the top, zoomed in on the total results (solid lines). The thickness of the lines is larger than the estimated error.}
\label{fig:NormCME_fixed_RhoR}
\end{figure}

\begin{figure}[hbt]
\centering
\includegraphics[width=0.45\textwidth]{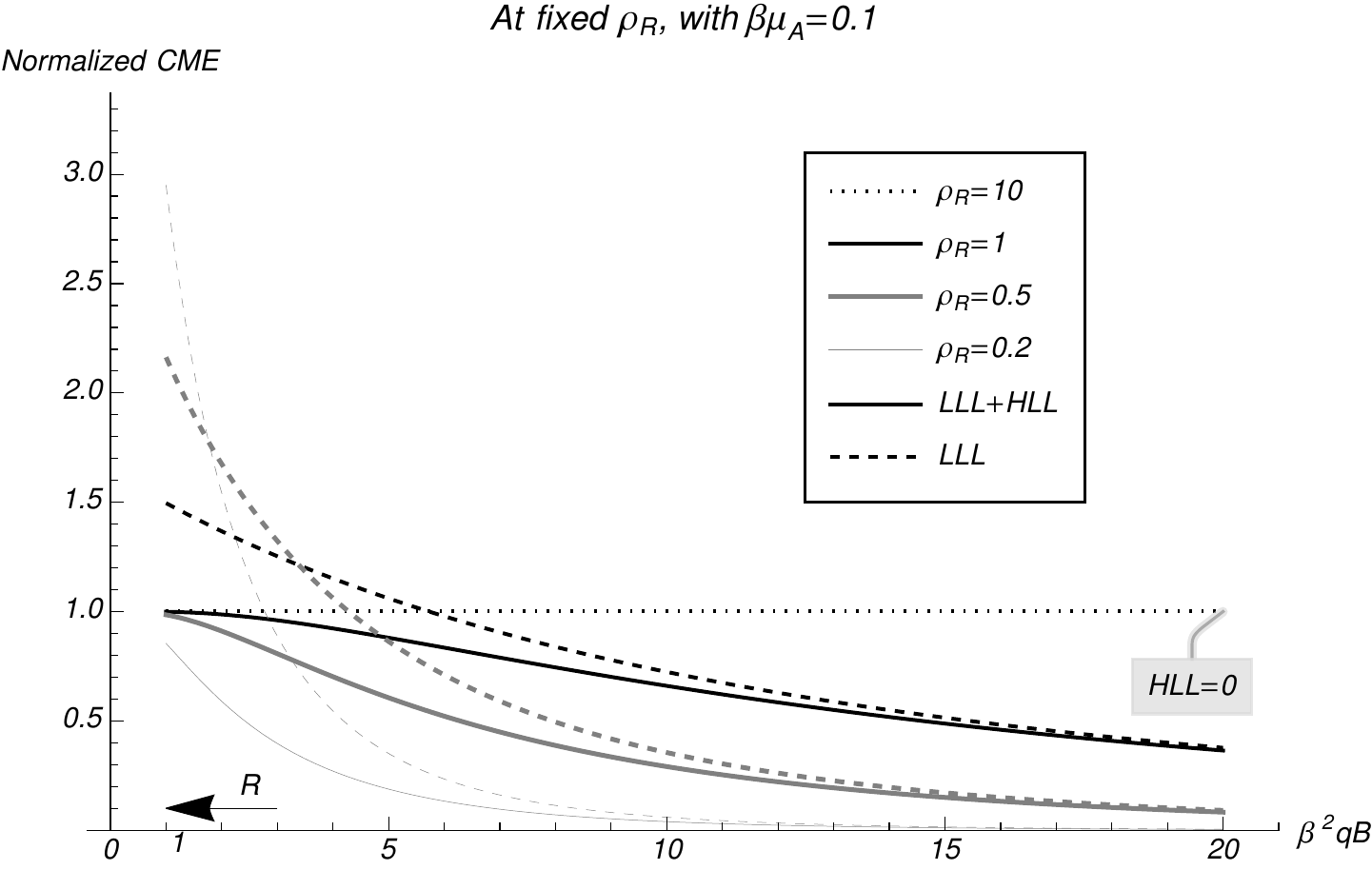}
\caption{Same as Fig.~\ref{fig:NormCME_fixed_RhoR} for a larger range of   $\beta^2|qB|$. }
\label{fig:NormCME_fixed_RhoR_2}
\end{figure}

We first report the value of the bound CME at the center of the cylinder $\rho=0$
where the CME reaches its peak. Figures \ref{fig:NormCME_fixed_RhoR} and \ref{fig:NormCME_fixed_RhoR_2}
show the normalized CME at the center of the cylinder obtained by fixing the parameter $\rho_R$
and varying the magnetic field intensity $|qB|$. Having fixed $\rho_R$, an increase in the
magnetic field corresponds to a decrease in the radius of the cylinder $R=\sqrt{2\rho_R/|qB|}$.
For small magnetic fields (large radius), the bound CME has the same value as the unbound one. However, as we commented earlier, its value is not completely
determined by the ground state, which can loosely be referred to as the LLL of the bound system. Contributions from higher ``Landau levels" (HLL)
are larger for smaller $\rho_R$; in Fig. \ref{fig:NormCME_fixed_RhoR_2} the HLL contributions
are negligible already at $\rho_R=10$. As we increase the magnetic field (decrease the radius)
the bound CME decreases and eventually vanishes. We also find that CME depends on
the temperature. If we  decrease the temperature while keeping the radius and
the magnetic field fixed, then $\beta^2|qB|$ increases and the CME is suppressed at very low temperatures.
Figures \ref{fig:NormCME_fixed_RhoR} and \ref{fig:NormCME_fixed_RhoR_2} do not reach the limit
$|qB|=0$ because lowering the magnetic field  becomes numerically expensive as it requires
the inclusions of higher Landau levels.

\begin{figure}[htb]
\centering
\includegraphics[width=0.45\textwidth]{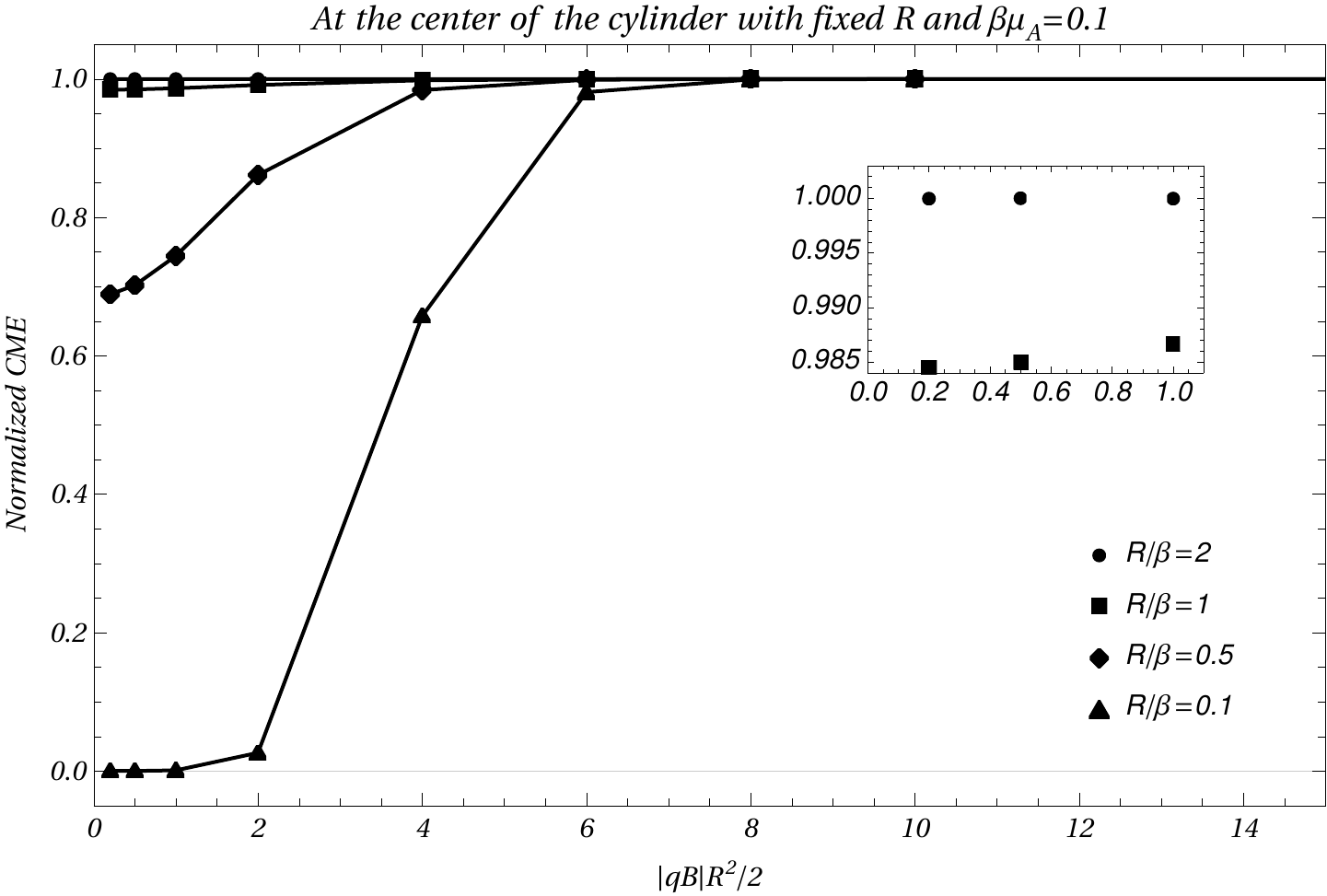}
\caption{The normalized CME as a function of the magnetic field keeping the radius of the cylinder $R$ constant.  The inset shows the details of the top-left part of the larger plot.}
\label{fig:NormCME_fixed_R}
\end{figure}

\begin{figure}[htb]
\centering
\includegraphics[width=0.45\textwidth]{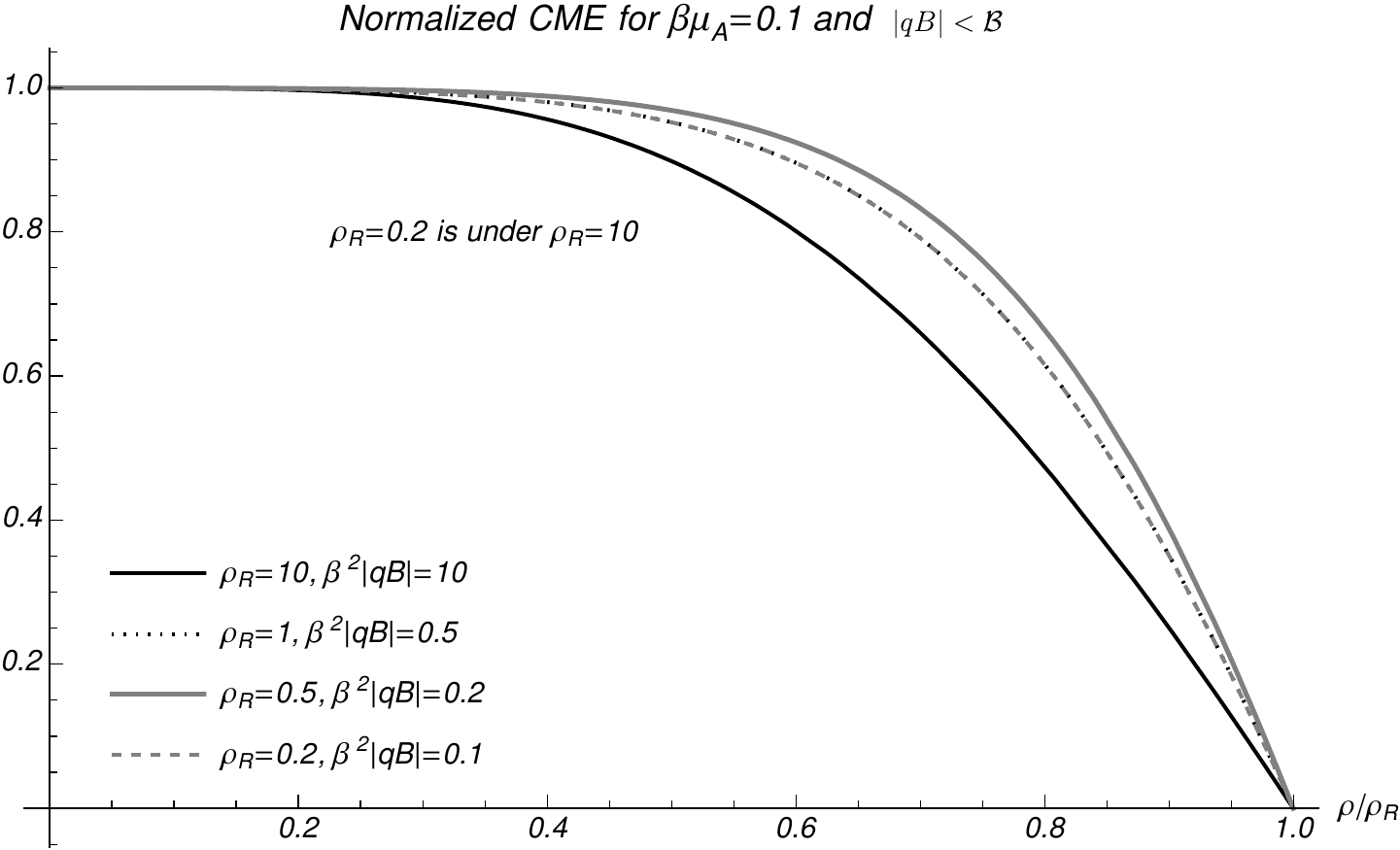}\\
\includegraphics[width=0.45\textwidth]{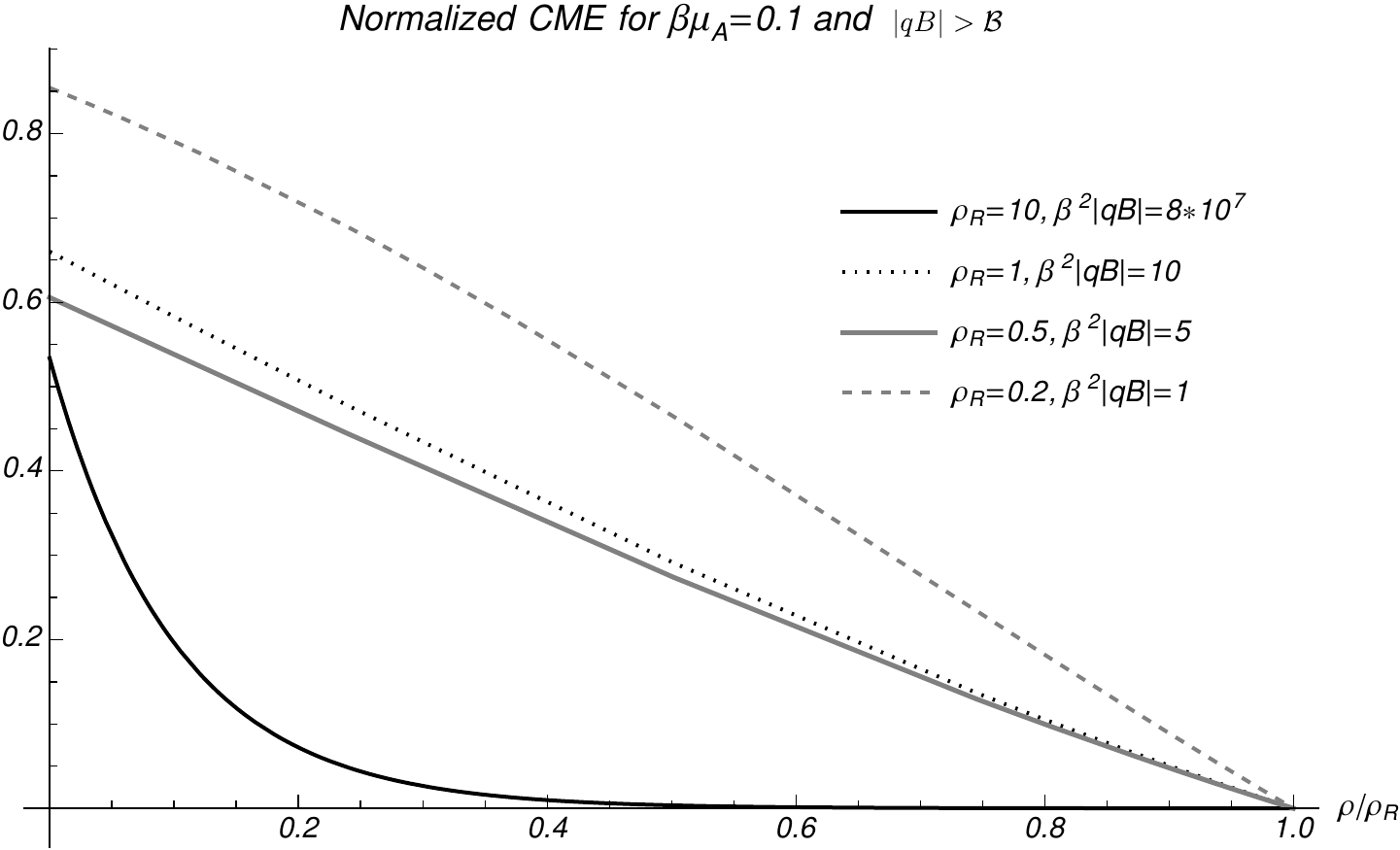}
\caption{The normalized CME as a function of the radial distance. The values of $|qB|$ are chosen with no particular criteria, and that is why the CME at the origin goes to different values for
$|qB|>\mathcal{B}$.}
\label{fig:CME_Radial}
\end{figure}

Let us now look at the CME in a common experimental setup: a fixed radius of the cylinder and a
varied magnetic field.  The results are shown in Fig. \ref{fig:NormCME_fixed_R}.
The bound CME approximately equals the unbound one in strong magnetic fields, whereas it is suppressed in weak fields. The parameter that distinguishes the two regimes is $\mathcal{B}(\rho_R)$ that we define below in (\ref{eq:B3}). The difference between the normalized CME at strong and weak magnetic fields is seen in Fig.~\ref{fig:CME_Radial} which displays the radial dependence of the effect. We can see that in both cases, the CME is larger at the center
of the cylinder and is vanishing at its surface  $\rho=\rho_R$. The difference between the two regimes is that the value of the normalized CME at the symmetry axis $\rho=0$ is close to unity for $|qB|<\mathcal{B}(\rho_R(|qB|,R))$ and is smaller than one otherwise. We also notice the different functional form of the $\rho$-dependence in the two cases.  

The vanishing of the CME current at the surface of the cylinder might be surprising at first glance because
it is not forbidden by the symmetry of the problem nor by the MIT boundary conditions.
Moreover, one might expect that the normal modes which are not present in the unbound case
are pushed toward the boundary of the cylinder and might cause an accumulation of chiral charge
similar to what is observed in the slab geometry~\cite{Valgushev:2015pjn}.
However, on closer inspection, we realize that the settings of the system forbid the
presence of a chiral charge at the surface. This can be seen evaluating the axial charge matrix element
using the solutions (\ref{eq:SolutionDiracFiniteR}), which reads
\begin{equation}
\begin{split}
j^0_A(\vec{x}|\lambda_k,\,m,\,p_z,\,\zeta) = &
\frac{qB}{2\pi^2}\left[C_1 C_3 F_1^2(\rho) + C_2 C_4 F_2^2(\rho) \right].
\end{split}
\end{equation}
At the boundary $\rho=\rho_R$, using (\ref{eq:MITF1F2}), the axial charge must 
vanish $j^0_A(R) =0$. That is why the CME always vanishes  at the surface of the cylinder.
The vanishing of the current at the boundary was also found for the chiral vortical effect
with MIT boundary conditions in the absence of a magnetic field \cite{Ambrus:2015lfr}.

\subsection{Qualitative analysis}
\label{sec:Boundary}

The effect of the boundary on  CME can be gauged by examining the contribution of the LLL, which is the level of the unbound system with $\lambda=n=0$. Indeed, we have already demonstrated that bound CME consists of contributions from the entire energy spectrum, whereas the unbound CME is driven exclusively by LLL
(see, for example, Fig.~\ref{fig:NormCME_fixed_RhoR}). As the cylinder radius decreases,  the LLL energy increases \cite{Buzzegoli:2022omv}. Additionally, the boundary condition lifts the degeneracy in the angular momentum
$m$. In our model the levels that converge onto the LLL of the unbound system are obtained from Eq. (\ref{eq:MITConstraint}). Among them is the lowest energy level with $m=1/2$, 
which we will call the LLL of a bound system. It scales with $\rho_R$ which in turn depends on $R$ and $B$. In particular, at large $\rho_R$, the corresponding principal quantum number is given by (\ref{eq:LambdaFirstRoot}).

The energy of LLL is given by (\ref{eq:EnergyFiniteCylinder}). Since $p_z\sim T$, the effect of the boundary on CME is essential only if $2|qB|\lambda_{\rm LLL}\beta$ is 
of order unity or larger. 
Define the characteristic value of the magnetic field as
\begin{equation}\label{eq:B3}
\beta^2\mathcal{B}(\rho_R) = \frac{1}{2\lambda_{\rm LLL}(\rho_R)}.
\end{equation}
The effect of the boundary is negligible if $|qB|\ll \mathcal{B}(\rho_R)$. We note that the characteristic value of the magnetic field is the increasing exponential function of $\rho_R$.

Although  (\ref{eq:B3}) reflects the particular boundary condition that we imposed, the qualitative picture is fairly general. That the energy levels of a system increase as its size decreases follows from the uncertainty relation. An increase in energy results in a suppression of thermal quantities, because
they are related to the Fermi-Dirac thermal distribution function.
The occupation number $n_F(\beta E)$ is small for high values of $\beta E$, leading to suppression of thermal quantities.
In the unbound case the lowest Landau level has the same energy independently of the
magnetic field because $\lambda$ is exactly zero. The CME conductivity is given completely
by the LLL and remains the same at any value of the magnetic field. Instead in the presence of
a boundary, we can always find a value of the magnetic field such that
$2\beta^2 \lambda_{\rm LLL}|qB|=1$.

\begin{figure}[htb]
\centering
\includegraphics[width=0.4\textwidth]{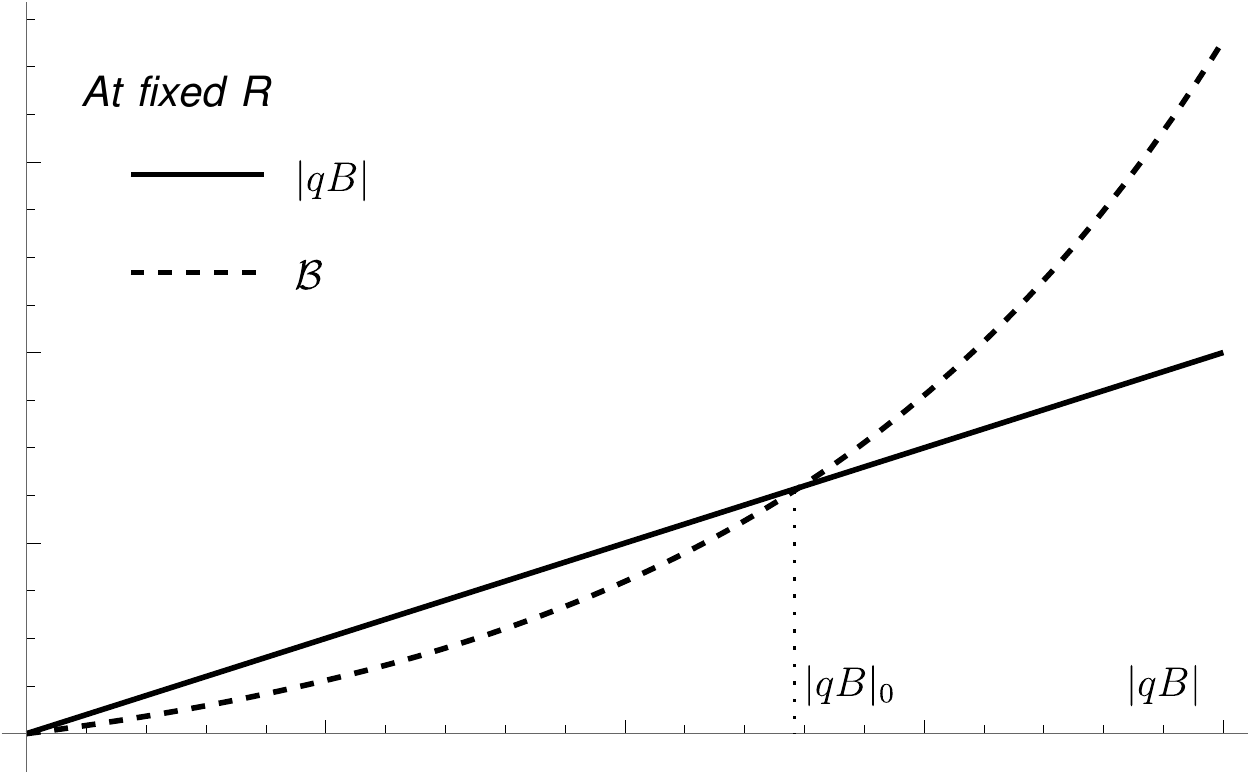}
\caption{The characteristic value of the magnetic field at a fixed radius of the cylinder $R$.}
\label{fig:CriticalB_fixed_R}
\end{figure}

We can now understand the results obtained for the bound CME.
If we keep $\rho_R$ fixed, then the characteristic value $\mathcal{B}(\rho_R)$ is also fixed.
Based on the above argument, when $|qB|$ is close to or larger than $\mathcal{B}$ the CME would be
suppressed. This is what we see in Fig. \ref{fig:NormCME_fixed_RhoR_2}.
Instead, if we fix the radius of the cylinder $R$, the value of $\rho_R$ changes with the
magnetic field, and so does the characteristic value $\mathcal{B}$. As shown in Fig.~\ref{fig:CriticalB_fixed_R},
if we fix the size of the cylinder, then $|qB|<\mathcal{B}(\rho_R)$ is always satisfied above a certain value
of the magnetic field $|qB|_0$. Namely, $|qB|>\mathcal{B}(\rho_R)$ when $|qB|<|qB|_0$ (weak field), and $|qB|<\mathcal{B}(\rho_R)$ when $|qB|>|qB|_0$ (strong field).

In the weak field (at fixed $R$), there is suppression of the CME, while in the strong field there is only minor effect as compared to the unbound case. This explains the results of Fig. \ref{fig:NormCME_fixed_R}. 

\begin{figure}[htb]
\centering
\includegraphics[width=\columnwidth]{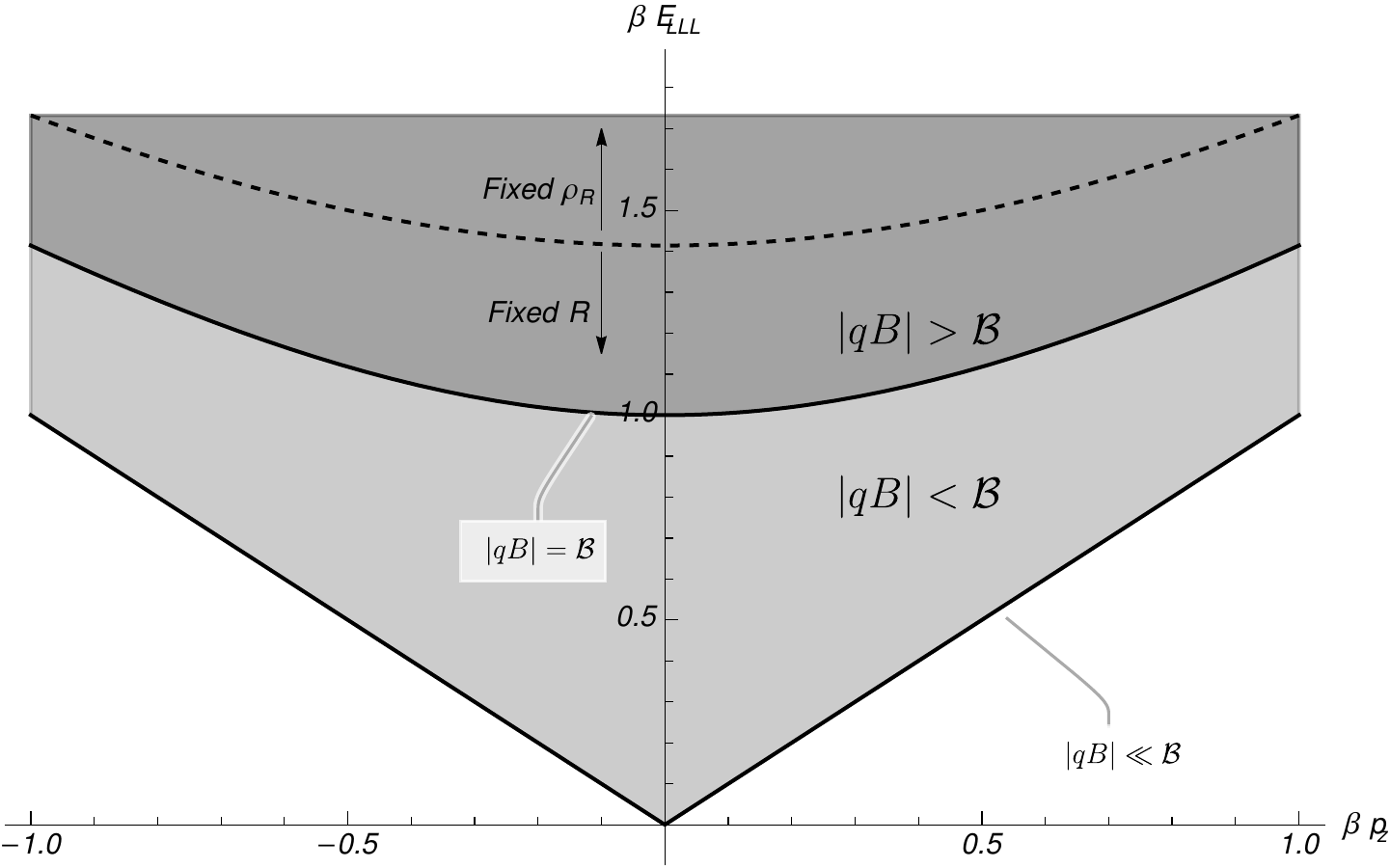}
\caption{The effect of the finite size of the cylinder on the LLL energy for different values of the magnetic field. When $|qB|<\mathcal{B}(\rho_R)$ the lowest Landau level lays
in the light shaded area and for $|qB|\ll\mathcal{B}(\rho_R)$ it  has negligible transverse momentum $2|qB|\lambda\approx 0$ (the lowest solid line). In the opposite limit 
$|qB|>\mathcal{B}(\rho_R)$ the LLL lays in the dark shaded area. The arrows
indicate the shift of the LLL energy if we increase the magnetic field at fixed $\rho_R$ or
at fixed $R$.}
\label{fig:Gapped_energy}
\end{figure}

What is described above is summarized in Fig.~\ref{fig:Gapped_energy} showing the dependence of the LLL energy on various  parameters. When $|qB|\ll\mathcal{B}(\rho_R)$, the transverse momentum
is negligible and the energy is only given by the longitudinal momentum $p_z$.
When $|qB|=\mathcal{B}(\rho_R)$, the LLL energy is $\beta E_{\rm LLL}=\sqrt{1+\beta^2 p_z^2}$.
For this energy and for energies above this, the effect of the boundary becomes strong and the
CME is suppressed. Starting at any values of $|qB|$ and $\rho_R$, if we increase the magnetic field
$|qB|$ keeping $\rho_R$ fixed, the LLL energy will increase resulting in further suppression
of the CME. Instead, if we increase $|qB|$ keeping the radius $R$ fixed we increase $\rho_R$, we
decrease the LLL energy and the CME will be less suppressed. The LLL energy is lowest for $\lambda=0$, i.e.\ in an unbound system.

\section{Conclusions}
\label{sec:Conclusions}
We studied a system of  free massless fermions confined in a cylinder with a
finite radius in the presence of a constant magnetic field using MIT boundary conditions.
The energy spectrum is different from the Landau levels obtained in an unbound space. The degeneracy with respect to the longitudinal total angular momentum $m$ is lifted due to the boundary conditions and the principal quantum number $\lambda$ is not an integer. The energy ground state of the bound system, which we also called LLL because it reduces to the ground state of the unbound system, depends on the system size. It increases when the cylinder radius $R$ decreases,  in agreement with the uncertainty relation. At finite $R$, the LLL does not have a definite chirality
and has a corresponding  principal quantum number given by Eq.~(\ref{eq:LambdaFirstRoot}) which tends to zero when $l_B \ll R$. In general, the effect of the boundary on the spectrum and by extension on all statistical quantities is important when the magnetic length  $l_B$ is of the order of or larger
than the radius of the cylinder $R$.

At finite temperature and finite chiral imbalance, we computed the CME.
The main result is shown in Fig. \ref{fig:NormCME_fixed_R}. For cylinders of the same size,
the CME is suppressed at very small magnetic fields and at very low temperatures.
The CME current is maximal at the center of the cylinder and it is vanishing at the boundary
of the cylinder. Even when the CME current is the same as in the unbound case, we found
that different from the unbound case where the CME is completely determined by the
lowest Landau level, in a finite region, all the levels contribute to the CME.

A suppression of the CME conductivity might seem incompatible with the non-renormalization 
of the chiral anomaly, which ultimately causes the CME. However, the anomaly only dictates that
the CME current must satisfy the relation (1) at the operator level, but does not prevent the
CME conductivity to be renormalized when the electric current operator is taken over a physical
state \cite{Anselm:1989gi,Adler:2004qt,Feng:2018tpb}. As a consequence, it is possible that the CME conductivity
differs when evaluated in bound states rather than unbound states as done in this work. 

It is important to stress that although we derived the result using thermal equilibrium with a conserved axial charge, the axial chemical potential is treated as a dynamical quantity. In our model $\mu_A$ only affects the thermal distribution functions but the ground state of the system is \emph{not} realized with a nonvanishing $\mu_A$. The use of thermal equilibrium is justified when looking at the system at timescales shorter than the relaxation time of the axial chemical potential. Therefore, a nonvanishing CME does not violate the conservation of energy as it does not describe a true steady state.
The finite-volume effects of the CME were also studied in \cite{Gorbar:2015wya,Valgushev:2015pjn,Sitenko:2016iqb} in a slab geometry at actual equilibrium with chiral imbalance using a model where the ground state \emph{does} have a fixed nonvanishing $\mu_A$. Accordingly, the total CME current must vanish in order to conserve energy \cite{Kharzeev:2023proc} in accordance with the Bloch theorem \cite{PhysRev.75.502}. This was confirmed by \cite{Gorbar:2015wya,Valgushev:2015pjn,Sitenko:2016iqb}. The complete absence of the CME current in \cite{Gorbar:2015wya,Sitenko:2016iqb} is then understood as a consequence of the boundary conditions explicitly preventing the flow of charges in that direction \cite{Gorbar:2015wya}, while the different geometry in \cite{Valgushev:2015pjn} allows for a local CME current.

In this paper we considered only one possible boundary condition.
Different conditions are likely to give different values near the boundary. However, as we argued above, the main cause of the CME suppression is that the LLL energy increases as we decrease the cylinder radius as dictated by the uncertainty principle. For this reason, it seems to us that our result should not significantly change if we impose different boundary conditions. For phenomenological assessments, it is important to show that this result is indeed not model dependent.
It would also be interesting to examine models with dynamical confinement instead of boundary conditions.

In practice, an accurate measurement of the boundary effects on CME may be possible in Dirac/Weyl semimetals. This requires $l_B>R$, where $R$ is the specimen linear size and the magnetic length is given by $l_B= 0.8 \mu\text{m}/\sqrt{B(\text{mT})}$. For example, for a semimetal with an observable CME in $B\sim$~mT magnetic fields, $R$ should be of the order of a $\mu$m.  In the relativistic heavy-ion collisions, $R\sim 10$~fm is the order of the system size, while $l_B$ can range from about a fm to pm at different collision energies. Therefore, the boundary effects are expected to be a significant factor in the CME phenomenology.

\bigskip
\acknowledgments
We are grateful to Semeon Valgushev for many fruitful discussions.
This work  was supported in part by the U.S. Department of Energy
under Grants No.\ DE-FG02-87ER40371 and No.\ DE-SC0023692.
%

\end{document}